\documentclass[12pt]{article}
\usepackage{amsmath,amssymb}
\usepackage{graphicx}
\newcommand{\eqdef}{\stackrel{\text{def}}{=}}
\newcommand{\n}{\nonumber\\}
\newcommand{\bm}{\boldsymbol}
\newcommand{\ignore}[1]{}
\numberwithin{equation}{section}
\newcommand{\Romannumeral}[1]{\uppercase\expandafter{\romannumeral#1}}
\newcommand{\I}{\text{\Romannumeral{1}}}
\newcommand{\II}{\text{\Romannumeral{2}}}

\allowdisplaybreaks[4]

\begin{document}

\baselineskip=20pt

\newfont{\elevenmib}{cmmib10 scaled\magstep1}
\newcommand{\preprint}{
   \begin{flushright}\normalsize \sf
     DPSU-13-4\\
   \end{flushright}}
\newcommand{\Title}[1]{{\baselineskip=26pt
   \begin{center} \Large \bf #1 \\ \ \\ \end{center}}}
\newcommand{\Author}{\begin{center}
   \large \bf Satoru Odake \end{center}}
\newcommand{\Address}{\begin{center}
     Department of Physics, Shinshu University,\\
     Matsumoto 390-8621, Japan
   \end{center}}
\newcommand{\Accepted}[1]{\begin{center}
   {\large \sf #1}\\ \vspace{1mm}{\small \sf Accepted for Publication}
   \end{center}}

\preprint
\thispagestyle{empty}

\Title{Equivalences of the Multi-Indexed\\ Orthogonal Polynomials}

\Author

\Address
\vspace{1cm}

\begin{abstract}
Multi-indexed orthogonal polynomials describe eigenfunctions of exactly
solvable shape-invariant quantum mechanical systems in one dimension
obtained by the method of virtual states deletion.
Multi-indexed orthogonal polynomials are labeled by a set of degrees
of polynomial parts of virtual state wavefunctions.
For multi-indexed orthogonal polynomials of Laguerre, Jacobi, Wilson
and Askey-Wilson types, two different index sets may give equivalent
multi-indexed orthogonal polynomials.
We clarify these equivalences. Multi-indexed orthogonal polynomials
with both type $\I$ and $\II$ indices are proportional to those of type
$\I$ indices only (or type $\II$ indices only) with shifted parameters.
\end{abstract}

\section{Introduction}
\label{intro}

Exactly solvable quantum mechanical systems in one dimension have seen
remarkable developments in recent years \cite{gomez}--\cite{os30}.
Important ingredients are new types of orthogonal polynomials; exceptional
orthogonal polynomials \cite{gomez,os16,os17,os23} and its generalization,
multi-indexed orthogonal polynomials \cite{gomez3,os25,os26,os27}.
Multi-indexed orthogonal polynomials are the main parts of eigenfunctions
of the shape-invariantly deformed quantum mechanical systems, which are
obtained from the original shape-invariant systems by the method of virtual
states deletion.
Multi-index orthogonal polynomials of Laguerre, Jacobi, Wilson,
Askey-Wilson, Racah and $q$-Racah types have been constructed.
They satisfy second order differential/difference equations and form
a complete set. Their degrees start from some positive integer $\ell$ and
Bochner's theorem \cite{bochner} and its generalization are avoided.

The method of virtual states deletion \cite{os25} is a systematic
procedure to obtain iso-spectral exactly solvable quantum mechanical systems
from the original exactly solvable systems.
It based on the Darboux-Crum transformation \cite{darb,crum} taking
virtual state wavefunctions as seed solutions.
The Darboux-Crum transformation in terms of eigenfunctions gives
Krein-Adler transformation \cite{adler}, by which some eigenstates of
the original system are deleted.
The Darboux-Crum transformation in terms of pseudo virtual state
wavefunctions \cite{os29,os30} gives a deformed system in which some extra
eigenstates are added to the original system.
Virtual and pseudo virtual state wavefunctions are obtained from
the eigenfunctions by using the discrete symmetries of the Hamiltonian
\cite{os25,os26,os27,os29,os30}.

Multi-indexed orthogonal polynomials are labeled by an index set
$\mathcal{D}=\{d_1,\ldots,d_M\}$, where $d_j$ is the degree of the polynomial
part of the virtual state wavefunction for deletion.
For Laguerre, Jacobi, Wilson and Askey-Wilson cases, we have two types
of virtual states, type $\I$ and type $\II$ \cite{os25,os27}.
For these cases, it may happen that the deformed system labeled by an
index set $\mathcal{D}$ and another deformed system labeled by a
different index set $\mathcal{D}'$ with shifted parameters are
equivalent, which means that the corresponding
two multi-indexed orthogonal polynomial labeled by $\mathcal{D}$ and
$\mathcal{D}'$ (with shifted parameters) are proportional.
In \cite{os25,os27}, we have presented such examples,
$\mathcal{D}=\{m^{\I},(m+1)^{\I},\ldots,(m+k-1)^{\I}\}$ and
$\mathcal{D}'=\{k^{\II},(k+1)^{\II},\ldots,(k+m-1)^{\II}\}$.
A certain extension of this result has been commented in \cite{os29,os30}, where
Wronskian/Casoratian identities and equivalence of two deformed systems
are obtained by using the Darboux-Crum transformation in terms of pseudo
virtual state wavefunctions.
In this paper we generalize this result.
By extending the Wronskian/Casoratian identities given in \cite{os29,os30},
we show that the multi-indexed orthogonal polynomials with both type $\I$
and $\II$ indices \eqref{D} are proportional to those with type $\I$ indices
only \eqref{PDnpropI} (or type $\II$ indices only \eqref{PDnpropII}).

This paper is organized as follows.
The Darboux-Crum approach to quantum mechanical systems and the multi-indexed
orthogonal polynomials is recapitulated in section \ref{sec:miop}.
After recalling the Wronskian/Casoratian identities, we extend them in
section \ref{sec:pvs}.
Section \ref{sec:equiv} is the main part of the paper.
By using the Wronskian/Casoratian identities obtained in \S\,\ref{sec:pvs},
we show the equivalence of two deformed systems obtained by the Darboux-Crum
transformation in terms of virtual state wavefunctions; one uses both type
$\I$ and $\II$ virtual states and the other uses type $\I$ only (or type
$\II$ only) with shifted parameters.
This implies that multi-indexed orthogonal polynomials with both type $\I$
and $\II$ indices are proportional to those with type $\I$ indices only
(or type $\II$ indices only) with shifted parameters.
The final section is for a summary and comments.
Various data of quantum mechanical systems described by Laguerre, Jacobi,
Wilson and Askey-Wilson polynomials are presented in Appendix.

\section{Quantum Mechanical Systems and Multi-Indexed\\ Orthogonal Polynomials}
\label{sec:miop}

Our analysis is based on the quantum mechanical formulation, in which
the Schr\"{o}dinger equation is a second order differential (oQM, ordinary
Quantum Mechanics) or difference (idQM, discrete Quantum Mechanics with
pure imaginary shifts).
See, for example, \cite{os24} for the general introduction.
Here we recapitulate the Darboux-Crum approach
to quantum mechanical systems with a continuous dynamical variable $x$
and the multi-indexed orthogonal polynomials
\cite{darb,crum,adler,os15,gos,os25,os27}.

We consider quantum mechanical systems in one dimension ($x_1<x<x_2$),
\begin{align}
  &\mathcal{H}=\mathcal{A}^{\dagger}\mathcal{A}\n
  &\phantom{\mathcal{H}}
  =\left\{
  \begin{array}{ll}
  p^2+U(x)&:\text{oQM}\\[2pt]
  \!\sqrt{V(x)}\,e^{\gamma p}\sqrt{V^*(x)}
  +\!\sqrt{V^*(x)}\,e^{-\gamma p}\sqrt{V(x)}-V(x)-V^*(x)
  &:\text{idQM}
  \end{array}\right.,\\
  &\mathcal{H}\phi_n(x)=\mathcal{E}_n\phi_n(x)\ \ (n\in\mathbb{Z}_{\geq 0}),
  \quad 0=\mathcal{E}_0<\mathcal{E}_1<\mathcal{E}_2<\cdots,\\
  &(\phi_n,\phi_m)\eqdef\int_{x_1}^{x_2}\!dx\,\phi_n(x)^*\phi_m(x)
  =h_n\delta_{nm}\ \ (0<h_n<\infty),
\end{align}
where the momentum operator is $p=-i\frac{d}{dx}$ and $\gamma$ is a real
number. The explicit forms of $\mathcal{A}$ and $\mathcal{A}^{\dagger}$
are given in Appendix.
Any solution of the Schr\"{o}dinger equation
$\mathcal{H}\tilde{\varphi}(x)=\tilde{\mathcal{E}}\tilde{\varphi}(x)$,
which needs not be square integrable, can be used as a seed solution for
the Darboux transformation.
The Hamiltonian $\mathcal{H}$ can be written as
\begin{equation}
  \mathcal{H}=\hat{\mathcal{A}}_{\tilde{\varphi}}^{\dagger}
  \hat{\mathcal{A}}_{\tilde{\varphi}}
  +\tilde{\mathcal{E}},\quad
  \hat{\mathcal{A}}_{\tilde{\varphi}}\tilde{\varphi}(x)=0,
\end{equation}
and the Darboux transformation is
\begin{equation}
  \mathcal{H}\to\mathcal{H}'=\hat{\mathcal{A}}_{\tilde{\varphi}}
  \hat{\mathcal{A}}_{\tilde{\varphi}}^{\dagger}
  +\tilde{\mathcal{E}},\quad
  \psi(x)\to\psi'(x)=\hat{\mathcal{A}}_{\tilde{\varphi}}\psi(x).
\end{equation}
Then it is easy to show
\begin{equation}
  \mathcal{H}\psi(x)=\mathcal{E}\psi(x)\Rightarrow
  \mathcal{H}'\psi'(x)=\mathcal{E}\psi'(x).
\end{equation}
Therefore we have $\mathcal{H}'\phi'_n(x)=\mathcal{E}_n\phi'_n(x)$.
The state corresponding to the used seed solution is deleted,
$\tilde{\varphi}'(x)=0$.
This procedure can be repeated.
Problems are (a) the deformed system $\mathcal{H}'$ is non-singular or not,
(b) there exist extra eigenstates other than $\phi'_n(x)$ or not.
Under the assumption that the deformed system is non-singular,
answer to (b) depends on the choice of seed solutions.
There are three cases:
(\romannumeral1)[state deleting] some eigenstates are deleted,
(\romannumeral2)[state adding] some eigenstates are added,
(\romannumeral3)[iso-spectral] no eigenstates are deleted or added.
We know several types of seed solutions, for example,
\begin{alignat}{3}
  &\tilde{\varphi}(x)=\phi_n(x)&&:\,\text{eigenstate}
  &&\Rightarrow\ \text{state deleting},\n
  &\tilde{\varphi}(x)=\tilde{\phi}^{\text{pvs}}_{\text{v}}(x)
  &&:\,\text{pseudo virtual state}
  &\ &\Rightarrow\ \text{state adding},\n
  &\tilde{\varphi}(x)=\tilde{\phi}^{\text{vs}}_{\text{v}}(x)
  &&:\,\text{virtual state}
  &&\Rightarrow\ \text{iso-spectral}.\nonumber
\end{alignat}
Virtual and pseudo virtual state wavefunctions are obtained from
eigenfunctions by using the discrete symmetry of the Hamiltonian.
For their definition, see \cite{os25,os27,os29,os30}.
After multi-step Darboux transformations, various quantities are expressed
in terms of the Wronskian $\text{W}[\cdots]$ or Casoratian
$\text{W}_{\gamma}[\cdots]$,
\begin{align}
  &\text{W}[f_1,f_2,\ldots,f_n](x)\eqdef
  \det\Bigl(\frac{d^{j-1}f_k(x)}{dx^{j-1}}\Bigr)_{1\le j,k\le n},\\
  &\text{W}_{\gamma}[f_1,f_2,\ldots,f_n](x)
  \eqdef i^{\frac12n(n-1)}
  \det\Bigl(f_k\bigl(x^{(n)}_j\bigr)\Bigr)_{1\leq j,k\leq n},\quad
  x_j^{(n)}\eqdef x+i(\tfrac{n+1}{2}-j)\gamma,
\end{align}
and $\text{W}[\cdot](x)=\text{W}_{\gamma}[\cdot](x)=1$ for $n=0$.
For example, after $s$-step (seed solutions
$\tilde{\varphi}_1(x),\ldots,$ $\tilde{\varphi}_s(x)$
are used in this order), eigenfunctions
$\phi^{[s]}_n(x)$ with energy $\mathcal{E}_n$ are
\begin{align}
  \text{oQM}:\ \ &\phi^{[s]}_n(x)=
  \frac{\text{W}[\tilde{\varphi}_1,\ldots,\tilde{\varphi}_s,\phi_n](x)}
  {\text{W}[\tilde{\varphi}_1,\ldots,\tilde{\varphi}_s](x)},
  \label{phisnoQM}\\
  \text{idQM}:\ \ &\phi^{[s]}_n(x)=\left(
  \frac{\sqrt{\prod_{j=0}^{s-1}V(x+i(\frac{s}{2}-j)\gamma)
  V^*(x-i(\frac{s}{2}-j)\gamma)}}
  {\text{W}_{\gamma}[\tilde{\varphi}_1,\ldots,\tilde{\varphi}_s]
  (x-i\frac{\gamma}{2})
  \text{W}_{\gamma}[\tilde{\varphi}_1,\ldots,\tilde{\varphi}_s]
  (x+i\frac{\gamma}{2})}\right)^{\frac12}\n
  &\phantom{\phi^{[s]}_n(x)=}
  \times
  \text{W}_{\gamma}[\tilde{\varphi}_1,\ldots,\tilde{\varphi}_s,\phi_n](x).
  \label{phisnidQM}
\end{align}
Multi-step Darboux transformation in terms of eigenfunctions only
(say case (A)) gives the Crum's theorem \cite{crum,os15} and its
generalization, Krein-Adler transformation \cite{adler,gos}.
That of pseudo virtual state wavefunctions only (say case (B)) leads to
Wronskian/Casoratian identities \cite{os29,os30}, which implies equivalence
of case (B) and case (A) with shifted parameters.
That of virtual state wavefunctions only  presents the multi-indexed
orthogonal polynomials \cite{os25,os27}.

The quantum systems to be considered have some parameters (coupling constants),
denoted symbolically by $\bm{\lambda}=(\lambda_1,\lambda_2,\ldots)$,
$\mathcal{H}=\mathcal{H}(\bm{\lambda})$,
$\mathcal{E}_n=\mathcal{E}_n(\bm{\lambda})$,
$\phi_n(x)=\phi_n(x;\bm{\lambda})$,
$P_n(\eta)=$ $P_n(\eta;\bm{\lambda})$,
etc.
Various data of quantum mechanical systems described by Laguerre(L),
Jacobi(J), Wilson(W) and Askey-Wilson(AW) polynomials are given in Appendix.

The multi-indexed orthogonal polynomials of L, J, W and AW types are the
main part of the eigenfunctions of the deformed systems obtained from the
original systems by the method of virtual states deletion.
There are two types of virtual states \cite{os25,os27},
type $\I$ $\tilde{\phi}^{\I}_{\text{v}}(x)$ and type $\II$
$\tilde{\phi}^{\II}_{\text{v}}(x)$ ($\text{v}\in\mathbb{Z}_{\geq 0}$),
which are derived by the discrete symmetries of the original Hamiltonian
and $\text{v}$ denotes the degree of polynomial part.
The set of virtual state wavefunctions for deletion is characterized by
their degrees:
\begin{align}
  &\mathcal{D}=\{d_1,\ldots,d_M\}=\{d^{\I}_1,\ldots,d^{\I}_{M_{\I}},
  d^{\II}_1,\ldots,d^{\II}_{M_{\II}}\}\quad
  (M_{\I},M_{\II}\geq 0,\ M=M_{\I}+M_{\II}),\n
  &\qquad(d^{\I}_j\in\mathbb{Z}_{\geq0},\text{ mutually distinct }
  ;\ d^{\II}_j\in\mathbb{Z}_{\geq0},\text{ mutually distinct}),
  \label{D}
\end{align}
namely we use seed solutions
\begin{equation*}
  \tilde{\phi}^{\I}_{d^{\I}_1}(x;\bm{\lambda}),\ldots,
  \tilde{\phi}^{\I}_{d^{\I}_{M_{\I}}}(x;\bm{\lambda}),
  \tilde{\phi}^{\II}_{d^{\II}_1}(x;\bm{\lambda}),\ldots,
  \tilde{\phi}^{\II}_{d^{\II}_{M_{\II}}}(x;\bm{\lambda}).
\end{equation*}
The deformed systems are \cite{os25,os27}
\begin{equation}
  \mathcal{H}_{\mathcal{D}}(\bm{\lambda})\phi_{\mathcal{D}\,n}(x;\bm{\lambda})
  =\mathcal{E}_n(\bm{\lambda})\phi_{\mathcal{D}\,n}(x;\bm{\lambda})\quad
  (n=0,1,\ldots).
\end{equation}
Here the Hamiltonians of the deformed systems are
\begin{align}
  \text{L,\,J}:\quad
  &\mathcal{H}_{\mathcal{D}}(\bm{\lambda})
  =p^2+U_{\mathcal{D}}(x;\bm{\lambda}),\quad
  U_{\mathcal{D}}(x;\bm{\lambda})
  =\frac{\partial_x^2\phi_{\mathcal{D}\,0}(x;\bm{\lambda})}
  {\phi_{\mathcal{D}\,0}(x;\bm{\lambda})},
  \label{UD}\\
  \text{W,\,AW}:\quad
  &\mathcal{H}_{\mathcal{D}}(\bm{\lambda})
  =\sqrt{V_{\mathcal{D}}(x;\bm{\lambda})}\,e^{\gamma p}
  \sqrt{V_{\mathcal{D}}^*(x;\bm{\lambda})}
  +\!\sqrt{V_{\mathcal{D}}^*(x;\bm{\lambda})}\,e^{-\gamma p}
  \sqrt{V_{\mathcal{D}}(x;\bm{\lambda})}\n
  &\qquad\qquad
  -V_{\mathcal{D}}(x;\bm{\lambda})-V_{\mathcal{D}}^*(x;\bm{\lambda}),\\
  &V_{\mathcal{D}}(x;\bm{\lambda})
  =V(x;\bm{\lambda}^{[M_{\I},M_{\II}]})\,
  \frac{\check{\Xi}_{\mathcal{D}}(x+i\frac{\gamma}{2};\bm{\lambda})}
  {\check{\Xi}_{\mathcal{D}}(x-i\frac{\gamma}{2};\bm{\lambda})}
  \frac{\check{\Xi}_{\mathcal{D}}(x-i\gamma;\bm{\lambda}+\bm{\delta})}
  {\check{\Xi}_{\mathcal{D}}(x;\bm{\lambda}+\bm{\delta})},
  \label{VD}
\end{align}
and their eigenfunctions are
\begin{align}
  &\phi_{\mathcal{D}\,n}(x;\bm{\lambda})\propto
  \psi_{\mathcal{D}}(x;\bm{\lambda})
  \check{P}_{\mathcal{D},n}(x;\bm{\lambda}),
  \label{phiDn}\\
  &\psi_{\mathcal{D}}(x;\bm{\lambda})
  =\phi_0(x;\bm{\lambda}^{[M_{\I},M_{\II}]})\times\left\{
  \begin{array}{ll}
  \check{\Xi}_{\mathcal{D}}(x;\bm{\lambda})^{-1}
  &:\text{L,\,J}\\
  \bigl(\check{\Xi}_{\mathcal{D}}(x-i\frac{\gamma}{2};\bm{\lambda})
  \,\check{\Xi}_{\mathcal{D}}(x+i\frac{\gamma}{2};\bm{\lambda})\bigr)^{-\frac12}
  &:\text{W,\,AW}
  \end{array}\right.,
  \label{psiDn}\\
  &\check{P}_{\mathcal{D},0}(x;\bm{\lambda})
  \propto\check{\Xi}_{\mathcal{D}}(x;\bm{\lambda}+\bm{\delta}),
  \label{PD0=XiD}
\end{align}
where $\bm{\lambda}^{[M_{\I},M_{\II}]}$ is
\begin{equation}
  \bm{\lambda}^{[M_{\I},M_{\II}]}
  \eqdef\bm{\lambda}+M_{\I}\tilde{\bm{\delta}}^{\I}
  +M_{\II}\tilde{\bm{\delta}}^{\II}.
  \label{laI_II}
\end{equation}
The shift $\bm{\delta}$ and twisted shifts $\tilde{\bm{\delta}}^{\I}$,
$\tilde{\bm{\delta}}^{\II}$ are given in Appendix for each polynomial.
The denominator polynomial $\check{\Xi}_{\mathcal{D}}(x;\bm{\lambda})\eqdef
\Xi_{\mathcal{D}}\bigl(\eta(x);\bm{\lambda}\bigr)$ and the multi-indexed
orthogonal polynomial $\check{P}_{\mathcal{D},n}(x;\bm{\lambda})\eqdef
P_{\mathcal{D},n}\bigl(\eta(x);\bm{\lambda}\bigr)$ are polynomials in
$\eta(x)$ and their degrees are generically\footnote{
For specific values of parameters, the coefficient of the highest power
term may vanish, see \S\,\ref{sec:A_coeff}.}
$\ell_{\mathcal{D}}$ and $\ell_{\mathcal{D}}+n$, respectively,
in which $\ell_{\mathcal{D}}$ is given by
\begin{align}
  \ell_{\mathcal{D}}
  &\eqdef\sum_{j=1}^{M_{\I}}d_j^{\I}-\frac12 M_{\I}(M_{\I}-1)
  +\sum_{j=1}^{M_{\II}}d_j^{\II}-\frac12M_{\II}(M_{\II}-1)+M_{\I}M_{\II}\n
  &=\sum_{j=1}^Md_j-\frac12M(M-1)+2M_{\I}M_{\II}.
  \label{ellD}
\end{align}
They are the polynomial parts of the Wronskian/Casoratian appearing in the
eigenfunctions \eqref{phisnoQM}--\eqref{phisnidQM}:
\begin{align}
  \check{\Xi}_{\mathcal{D}}(x;\bm{\lambda})&\propto
  \text{polynomial part of }
  \text{W}_{\gamma}[\tilde{\phi}^{\I}_{d^{\I}_1},\ldots,
  \tilde{\phi}^{\I}_{d^{\I}_{M_{\I}}},
  \tilde{\phi}^{\II}_{d^{\II}_1},\ldots,
  \tilde{\phi}^{\II}_{d^{\II}_{M_{\II}}}]
  (x;\bm{\lambda}),\\
  \check{P}_{\mathcal{D},n}(x;\bm{\lambda})&\propto
  \text{polynomial part of }
  \text{W}_{\gamma}[\tilde{\phi}^{\I}_{d^{\I}_1},\ldots,
  \tilde{\phi}^{\I}_{d^{\I}_{M_{\I}}},
  \tilde{\phi}^{\II}_{d^{\II}_1},\ldots,
  \tilde{\phi}^{\II}_{d^{\II}_{M_{\II}}},
  \phi_n](x;\bm{\lambda}),
\end{align}
for W and AW cases ($\text{W}_{\gamma}[\cdots]$ are replaced by
$\text{W}[\cdots]$ for L and J cases).
We remark that the deformed Hamiltonian
$\mathcal{H}_{\mathcal{D}}(\bm{\lambda})$ is independent of the order of
$d_j$ but the denominator polynomial $\Xi_{\mathcal{D}}(\eta;\bm{\lambda})$
and the multi-indexed orthogonal polynomial
$P_{\mathcal{D},n}(\eta;\bm{\lambda})$ may change sign under the permutation
of $d_j$'s (The set $\mathcal{D}$ of $\Xi_{\mathcal{D}}(\eta;\bm{\lambda})$
and $P_{\mathcal{D},n}(\eta;\bm{\lambda})$ is understood as an ordered set).
The explicit forms of $\Xi_{\mathcal{D}}(\eta;\bm{\lambda})$ and
$P_{\mathcal{D},n}(\eta;\bm{\lambda})$ are given in Appendix.

The deformed Hamiltonian $\mathcal{H}_{\mathcal{D}}(\bm{\lambda})$
is non-singular for large enough $g$ (L), large enough $g$ and $h$ (J),
large enough $\text{Re}\,a_j$ ($\forall j$) (W) and small enough $|a_j|$
($\forall j$) (AW). For more details, see \cite{os25,os27}.

\section{Pseudo Virtual States and Wronskian/Casoratian\\ Identities}
\label{sec:pvs}

By using pseudo virtual states which are obtained from eigenstate by using
the discrete symmetry of the original Hamiltonian, some Wronskian identities
\cite{os29} and Casoratian identities \cite{os30} are derived.
These identities imply the equivalence of two deformed quantum systems
obtained by Darboux-Crum transformation in terms of pseudo virtual
state wavefunctions and eigenfunctions with shifted parameters.
Here we extend these results.
The data of pseudo virtual states are given in Appendix.

First we consider W and AW cases.
In \cite{os30} the followings are shown.
Let a set $\mathcal{D}$ and an integer $N$ be
\begin{equation}
  \mathcal{D}=\{d_1,\ldots,d_M\}
  \ \ (d_j\in\mathbb{Z}_{\geq 0},\ \text{mutually distinct}),\quad
  N\geq\max\{d_1,\ldots,d_M\}.
\end{equation}
A set $\bar{\mathcal{D}}$ and $\bar{\bm{\lambda}}$ are defined by
\begin{align}
  &\bar{\mathcal{D}}\eqdef\{0,1,\ldots,N\}\backslash\{N-d_1,\ldots,N-d_M\}
  \eqdef\{e_1,\ldots,e_{N+1-M}\},\\
  &\bar{\bm{\lambda}}\eqdef\bm{\lambda}-(N+1)\bm{\delta}.
\end{align}
Deform the original systems $\mathcal{H}(\bm{\lambda})$ and
$\mathcal{H}(\bar{\bm{\lambda}})$ by Darboux-Crum transformation
with the following seed solutions:
\begin{alignat*}{2}
  \mathcal{H}(\bm{\lambda})\ \ \text{with}&
  \ \ \tilde{\phi}^{\text{pvs}}_{d_1}(x;\bm{\lambda}),\ldots,
  \tilde{\phi}^{\text{pvs}}_{d_M}(x;\bm{\lambda})
  &\ \Rightarrow\ &\mathcal{H}^{\text{pvs}}_{\mathcal{D}}(\bm{\lambda}),\\
  \mathcal{H}(\bar{\bm{\lambda}})\ \ \text{with}&
  \ \ \phi_{e_1}(x;\bar{\bm{\lambda}}),\ldots,
  \phi_{e_{N+1-M}}(x;\bar{\bm{\lambda}})
  &\ \Rightarrow\ &\mathcal{H}^{\text{KA}}_{\bar{\mathcal{D}}}
  (\bar{\bm{\lambda}}).
\end{alignat*}
The denominator polynomials $\Xi^{\text{pvs}}_{\mathcal{D}}(\eta;\bm{\lambda})$
and $\bar{\Xi}_{\bar{\mathcal{D}}}(\eta;\bar{\bm{\lambda}})$ are given by
\begin{align}
  \Xi_{\mathcal{D}}^{\text{pvs}}\bigl(\eta(x);\bm{\lambda}\bigr)
  &\propto\text{polynomial part of }
  \text{W}_{\gamma}[\tilde{\phi}^{\text{pvs}}_{d_1},\ldots,
  \tilde{\phi}^{\text{pvs}}_{d_M}](x;\bm{\lambda}),\\
  \bar{\Xi}_{\bar{\mathcal{D}}}\bigl(\eta(x);\bar{\bm{\lambda}}\bigr)
  &\propto\text{polynomial part of }
  \text{W}_{\gamma}[\phi_{e_1},\ldots,\phi_{e_{N+1-M}}](x;\bar{\bm{\lambda}}).
\end{align}
Then we have the Casoratian identity
\begin{equation}
  \Xi^{\text{pvs}}_{\mathcal{D}}(\eta;\bm{\lambda})
  \propto\bar{\Xi}_{\bar{\mathcal{D}}}(\eta;\bar{\bm{\lambda}}),
  \label{Xiid1}
\end{equation}
and this implies the equivalence of two deformed systems
\begin{equation}
  \mathcal{H}^{\text{pvs}}_{\mathcal{D}}(\bm{\lambda})
  -\mathcal{E}_{-N-1}(\bm{\lambda})
  =\kappa^{-N-1}\mathcal{H}^{\text{KA}}_{\bar{\mathcal{D}}}
  (\bar{\bm{\lambda}}).
  \label{Heq1}
\end{equation}
These Hamiltonians are non-singular if the index set $\bar{\mathcal{D}}$
satisfies Krein-Adler (KA) condition, $\prod_{j=1}^{N+1-M}(n-e_j)\geq 0$
($\forall n\in\mathbb{Z}_{\geq 0}$) \cite{adler,gos}.
These relations are based on the energy relation
\begin{equation}
  \tilde{\mathcal{E}}^{\text{pvs}}_{\text{v}}(\bm{\lambda})
  -\mathcal{E}_{-N-1}(\bm{\lambda})
  =\kappa^{-N-1}
  \mathcal{E}_{N-\text{v}}(\bar{\bm{\lambda}}).
\end{equation}
The choice of the integer $N$ is not unique and the systems with different
$N$ are related by shape-invariance.

The spectrum of $\mathcal{H}^{\text{pvs}}_{\mathcal{D}}(\bm{\lambda})$ and
$\mathcal{H}^{\text{KA}}_{\bar{\mathcal{D}}}(\bar{\bm{\lambda}})$ are
\begin{align*}
  \mathcal{H}^{\text{pvs}}_{\mathcal{D}}(\bm{\lambda})
  &:\ \ \tilde{\mathcal{E}}^{\text{pvs}}_{\text{v}}(\bm{\lambda})
  \ (\text{v}\in\mathcal{D}),
  \ \ \mathcal{E}_n(\bm{\lambda})\ (n\in\mathbb{Z}_{\geq 0}),\\
  \mathcal{H}^{\text{KA}}_{\bar{\mathcal{D}}}(\bar{\bm{\lambda}})
  &:\ \ \mathcal{E}_{N-\text{v}}(\bar{\bm{\lambda}})
  \ (\text{v}\in\mathcal{D}),
  \ \ \mathcal{E}_{N+1+n}(\bar{\bm{\lambda}})
  \ (n\in\mathbb{Z}_{\geq 0}),
\end{align*}
and the energy eigenvalues satisfy
\begin{equation}
  \mathcal{E}_n(\bm{\lambda})-\mathcal{E}_{-N-1}(\bm{\lambda})
  =\kappa^{-N-1}\mathcal{E}_{N+1+n}(\bar{\bm{\lambda}}).
\end{equation}
Therefore we can further perform Darboux-Crum transformation taking these
eigenfunctions with energy $\mathcal{E}_n(\bm{\lambda})$ and
$\mathcal{E}_{N+1+n}(\bar{\bm{\lambda}})$ as seed solutions,
which originate from $\phi_n(x;\bm{\lambda})$ and
$\phi_{N+1+n}(x;\bar{\bm{\lambda}})$,
for $\mathcal{H}^{\text{pvs}}_{\mathcal{D}}(\bm{\lambda})$ and
$\mathcal{H}^{\text{KA}}_{\bar{\mathcal{D}}}(\bar{\bm{\lambda}})$ respectively.
Let the set of $M'$ integers for this extra deletion be
$\{d'_1,\ldots,d'_{M'}\}$ ($d'_j\in\mathbb{Z}_{\geq 0}$, mutually distinct).
Then we have
\begin{alignat*}{2}
  \mathcal{H}^{\text{pvs}}_{\mathcal{D}}(\bm{\lambda})\ \ \text{with}&
  \ \ \phi^{\text{pvs}}_{\mathcal{D}\,d'_1}(x;\bm{\lambda}),\ldots,
  \phi^{\text{pvs}}_{\mathcal{D}\,d'_{M'}}(x;\bm{\lambda})
  &\ \Rightarrow\ &\mathcal{H}^{\text{pvs+es}}_{\mathcal{D}_{\text{ext}}}
  (\bm{\lambda}),\\
  \mathcal{H}^{\text{KA}}_{\bar{\mathcal{D}}}(\bar{\bm{\lambda}})
  \ \ \text{with}&
  \ \ \phi^{\text{KA}}_{\bar{\mathcal{D}}\,N+1+d'_1}(x;\bar{\bm{\lambda}}),
  \ldots,
  \phi^{\text{KA}}_{\bar{\mathcal{D}}\,N+1+d'_{M'}}(x;\bar{\bm{\lambda}})
  &\ \Rightarrow\ &\mathcal{H}^{\text{KA}}_{\bar{\mathcal{D}}_{\text{ext}}}
  (\bar{\bm{\lambda}}),
\end{alignat*}
namely
\begin{align*}
  \mathcal{H}(\bm{\lambda})\ \ \text{with}&
  \ \ \tilde{\phi}^{\text{pvs}}_{d_1}(x;\bm{\lambda}),\ldots,
  \tilde{\phi}^{\text{pvs}}_{d_M}(x;\bm{\lambda}),
  \phi_{d'_1}(x;\bm{\lambda}),\ldots,
  \phi_{d'_{M'}}(x;\bm{\lambda})
  \ \Rightarrow\ \mathcal{H}^{\text{pvs+es}}_{\mathcal{D}_{\text{ext}}}
  (\bm{\lambda}),\\
  \mathcal{H}(\bar{\bm{\lambda}})\ \ \text{with}&
  \ \ \phi_{e_1}(x;\bar{\bm{\lambda}}),\ldots,
  \phi_{e_{N+1-M}}(x;\bar{\bm{\lambda}}),
  \phi_{N+1+d'_1}(x;\bar{\bm{\lambda}}),\ldots,
  \phi_{N+1+d'_{M'}}(x;\bar{\bm{\lambda}})
  \ \Rightarrow\ \mathcal{H}^{\text{KA}}_{\bar{\mathcal{D}}_{\text{ext}}}
  (\bar{\bm{\lambda}}),
\end{align*}
where
\begin{align}
  &\mathcal{D}_{\text{ext}}=\{d_1,\ldots,d_M,d'_1,\ldots,d'_{M'}\},\\
  &\bar{\mathcal{D}}_{\text{ext}}
  =\{e_1,\ldots,e_{N+1-M},N+1+d'_1,\ldots,N+1+d'_{M'}\}.
\end{align}
Since the same procedure is applied to equivalent systems, we have
\begin{equation}
  \mathcal{H}^{\text{pvs+es}}_{\mathcal{D}_{\text{ext}}}(\bm{\lambda})
  -\mathcal{E}_{-N-1}(\bm{\lambda})
  =\kappa^{-N-1}\mathcal{H}^{\text{KA}}_{\bar{\mathcal{D}}_{\text{ext}}}
  (\bar{\bm{\lambda}}).
  \label{Heq2}
\end{equation}
These Hamiltonians are non-singular if $\bar{\mathcal{D}}_{\text{ext}}$
satisfies KA condition.
By defining $\Xi^{\text{pvs+es}}_{\mathcal{D}_{\text{ext}}}
(\eta;\bm{\lambda})$ and 
$\bar{\Xi}_{\bar{\mathcal{D}}_{\text{ext}}}(\eta;\bar{\bm{\lambda}})$ as
\begin{align}
  \Xi^{\text{pvs+es}}_{\mathcal{D}_{\text{ext}}}
  \bigl(\eta(x);\bm{\lambda}\bigr)&\propto
  \text{polynomial part of }
  \text{W}_{\gamma}[\tilde{\phi}^{\text{pvs}}_{d_1},\ldots,
  \tilde{\phi}^{\text{pvs}}_{d_M},
  \phi_{d'_1},\ldots,\phi_{d'_{M'}}](x;\bm{\lambda}),\\
  \bar{\Xi}_{\bar{\mathcal{D}}_{\text{ext}}}
  \bigl(\eta(x);\bar{\bm{\lambda}}\bigr)&\propto
  \text{polynomial part of }\n
  &\qquad
  \text{W}_{\gamma}[\phi_{e_1},\ldots,\phi_{e_{N+1-M}},
  \phi_{N+1+d'_1},\ldots,\phi_{N+1+d'_{M'}}](x;\bar{\bm{\lambda}}),
\end{align}
we obtain
\begin{equation}
  \Xi^{\text{pvs+es}}_{\mathcal{D}_{\text{ext}}}(\eta;\bm{\lambda})
  \propto\bar{\Xi}_{\bar{\mathcal{D}}_{\text{ext}}}
  (\eta;\bar{\bm{\lambda}}).
  \label{Xiid2}
\end{equation}
This polynomial proportionality \eqref{Xiid2} and equality \eqref{Heq2}
as an algebraic equation are valid for any parameter ranges (except for
the specific values mentioned in footnote 1).

For L and J cases, the above argument is valid by replacing Casoratian
$\text{W}_{\gamma}[\cdots]$ with Wronskian $\text{W}[\cdots]$ and
setting $\kappa=1$.
Eqs.\eqref{Xiid1} and \eqref{Heq1} are shown in \cite{os29}.
Further Darboux-Crum transformation in terms of eigenfunctions gives
\eqref{Heq2} and \eqref{Xiid2}.

\section{Equivalence of Multi-Indexed Orthogonal Polynomials}
\label{sec:equiv}

We will show that the deformed system obtained by the Darboux-Crum
transformations in terms of both type $\I$ and $\II$ virtual state
wavefunctions is equivalent to that of type $\I$ only (or type $\II$
only) with shifted parameters.
This means that multi-indexed orthogonal polynomials with both type $\I$
and $\II$ indices are proportional to that of type $\I$ only (or type $\II$
only) with shifted parameters.
Our starting point is the proportionality of denominator polynomials
\eqref{Xiid2}, namely (changing the notation :
$(d_j,d'_j,M,M')\to(d'_j,d''_j,M',M'')$.
$N\geq\max\{d'_1,\ldots,d'_{M'}\}$),
\begin{align}
  &\text{polynomial part of }
  \text{W}_{\gamma}[\tilde{\phi}^{\text{pvs}}_{d'_1},\ldots,
  \tilde{\phi}^{\text{pvs}}_{d'_{M'}},
  \phi_{d''_1},\ldots,\phi_{d''_{M''}}](x;\bm{\lambda})
  \label{Xiid3}\\
  \propto\ &\text{polynomial part of }
  \text{W}_{\gamma}[\phi_{e_1},\ldots,\phi_{e_{N+1-M'}},
  \phi_{N+1+d''_1},\ldots,\phi_{N+1+d''_{M''}}]
  \bigl(x;\bm{\lambda}-(N+1)\bm{\delta}\bigr),\nonumber
\end{align}
or its Wronskian version.
This relation is valid for any parameter ranges (except for specific
parameter values mentioned in footnote 1).
Various data are given in Appendix.

\subsection{Wilson and Askey-Wilson}
\label{sec:equiv_WAW}

First we consider W and AW cases.
{}From the relation between (pseudo) virtual state and eigenstate
\begin{equation*}
  \tilde{\phi}^{\I}_{\text{v}}(x;\bm{\lambda})
  =\phi_{\text{v}}\bigl(x;\mathfrak{t}^{\I}(\bm{\lambda})\bigl),\quad
  \tilde{\phi}^{\II}_{\text{v}}(x;\bm{\lambda})
  =\phi_{\text{v}}\bigl(x;\mathfrak{t}^{\II}(\bm{\lambda})\bigl),\quad
  \tilde{\phi}^{\text{pvs}}_{\text{v}}(x;\bm{\lambda})
  =\phi_{\text{v}}\bigl(x;\mathfrak{t}(\bm{\lambda})\bigl),
\end{equation*}
and $\mathfrak{t}=\mathfrak{t}^{\II}\circ\mathfrak{t}^{\I}$, we have
\begin{equation}
  \tilde{\phi}^{\text{pvs}}_{\text{v}}
  \bigl(x;\mathfrak{t}^{\I}(\bm{\lambda})\bigr)
  =\tilde{\phi}^{\II}_{\text{v}}(x;\bm{\lambda}),\quad
  \tilde{\phi}^{\text{pvs}}_{\text{v}}
  \bigl(x;\mathfrak{t}^{\II}(\bm{\lambda})\bigr)
  =\tilde{\phi}^{\I}_{\text{v}}(x;\bm{\lambda}).
\end{equation}
For the index set $\mathcal{D}$ \eqref{D}, we take an integer $N$ and
define an index set $\bar{\mathcal{D}}_{\I}$
\begin{align}
  &N\geq d^{\text{max}}_{\II}\eqdef\max\{d^{\II}_1,\ldots,d^{\II}_{M_{\II}}\}
  \ \ (\max\{\cdot\}\eqdef-1),\\
  &\bar{\mathcal{D}}_{\I}\eqdef\Bigl(\{0,1,\ldots,N\}\backslash
  \{N-d^{\II}_1,\ldots,N-d^{\II}_{M_{\II}}\}\Bigr)
  \cup\{N+1+d^{\I}_1,\ldots,N+1+d^{\I}_{M_{\I}}\}\n
  &\phantom{\bar{\mathcal{D}}_{\I}}
  \eqdef\{e^{\I}_1,e^{\I}_2,\ldots,e^{\I}_{N+1-M_{\II}+M_{\I}}\}.
  \label{DbI}
\end{align}
Since \eqref{Xiid3} holds for any parameter ranges, let us substitute
$\bm{\lambda}\to\mathfrak{t}^{\I}(\bm{\lambda})$ in \eqref{Xiid3}. 
Then we have
\begin{align}
  &\text{polynomial part of }
  \text{W}_{\gamma}[\tilde{\phi}^{\II}_{d^{\II}_1},\ldots,
  \tilde{\phi}^{\II}_{d^{\II}_{M_{\II}}},
  \tilde{\phi}^{\I}_{d^{\I}_1},\ldots,\tilde{\phi}^{\I}_{d^{\I}_{M_{\I}}}]
  (x;\bm{\lambda})\n
  \propto\ &\text{polynomial part of }
  \text{W}_{\gamma}[\tilde{\phi}^{\I}_{e^{\I}_1},\ldots,
  \tilde{\phi}^{\I}_{e^{\I}_{N+1-M_{\II}+M_{\I}}}]
  \bigl(x;\bm{\lambda}-(N+1)\tilde{\bm{\delta}}^{\I}\bigr),
\end{align}
where $\mathfrak{t}^{\I}+\beta\bm{\delta}
=\mathfrak{t}^{\I}(\bm{\lambda}+\beta\tilde{\bm{\delta}}^{\I})$ is used.
Therefore the denominator polynomial with both type $\I$ and $\II$ indices
is proportional to one of type $\I$ only,
\begin{equation}
  \Xi_{\mathcal{D}}(\eta;\bm{\lambda})\propto
  \Xi_{\bar{\mathcal{D}}_{\I}}\bigl(\eta;\bm{\lambda}
  -(N+1)\tilde{\bm{\delta}}^{\I}\bigr).
  \label{XiDpropI}
\end{equation}
The relation $\tilde{\bm{\delta}}^{\II}=-\tilde{\bm{\delta}}^{\I}$ implies
\begin{equation}
  \bm{\lambda}-(N+1)\tilde{\bm{\delta}}^{\I}
  +(N+1-M_{\II}+M_{\I})\tilde{\bm{\delta}}^{\I}
  =\bm{\lambda}^{[M_{\I},M_{\II}]}.
  \label{la-(N+1)I}
\end{equation}
By using this, \eqref{XiDpropI} and the general formula for potential function
\eqref{VD}, we can show that
\begin{equation}
  V_{\mathcal{D}}(x;\bm{\lambda})
  =V_{\bar{\mathcal{D}}_{\I}}
  \bigl(x;\bm{\lambda}-(N+1)\tilde{\bm{\delta}}^{\I}\bigr).
\end{equation}
Namely we obtain the equivalence of two deformed systems,
type I and II mixed system and type I system with shifted parameter,
\begin{equation}
  \mathcal{H}_{\mathcal{D}}(\bm{\lambda})=\mathcal{H}_{\bar{\mathcal{D}}_{\I}}
  \bigl(\bm{\lambda}-(N+1)\tilde{\bm{\delta}}^{\I}\bigr).
  \label{HD=HDbI}
\end{equation}
This equality as an algebraic equation holds for any parameter ranges
but this deformed Hamiltonian is non-singular only for restricted
parameter ranges (see the end of \S\,\ref{sec:miop}).
{}From the general formula \eqref{psiDn}, we can show
\begin{equation}
  \psi_{\mathcal{D}}(x;\bm{\lambda})\propto\psi_{\bar{\mathcal{D}}_{\I}}
  \bigl(x;\bm{\lambda}-(N+1)\tilde{\bm{\delta}}^{\I}\bigr).
  \label{psiDprop}
\end{equation}
Therefore the general formula for eigenfunctions \eqref{phiDn} gives
\begin{equation}
  P_{\mathcal{D},n}(\eta;\bm{\lambda})\propto
  P_{\bar{\mathcal{D}}_{\I},n}\bigl(\eta;\bm{\lambda}
  -(N+1)\tilde{\bm{\delta}}^{\I}\bigr)\quad(n=0,1,\ldots).
  \label{PDnpropI}
\end{equation}
This proportionality is valid for any parameter ranges.

Similarly, by substituting $\bm{\lambda}\to\mathfrak{t}^{\II}(\bm{\lambda})$
in \eqref{Xiid3}, we can show that a type $\I$ and $\II$ mixed system is
equivalent to a type $\II$ system with shifted parameters.
Results are the following:
\begin{align}
  &N\geq d^{\text{max}}_{\I}\eqdef\max\{d^{\I}_1,\ldots,d^{\I}_{M_{\I}}\},
  \label{NII}\\
  &\bar{\mathcal{D}}_{\II}\eqdef\Bigl(\{0,1,\ldots,N\}\backslash
  \{N-d^{\I}_1,\ldots,N-d^{\I}_{M_{\I}}\}\Bigr)
  \cup\{N+1+d^{\II}_1,\ldots,N+1+d^{\II}_{M_{\II}}\}\n
  &\phantom{\bar{\mathcal{D}}_{\II}}
  \eqdef\{e^{\II}_1,e^{\II}_2,\ldots,e^{\II}_{N+1-M_{\I}+M_{\II}}\},
  \label{DbII}\\
  &\Xi_{\mathcal{D}}(\eta;\bm{\lambda})\propto
  \Xi_{\bar{\mathcal{D}}_{\II}}\bigl(\eta;\bm{\lambda}
  -(N+1)\tilde{\bm{\delta}}^{\II}\bigr),
  \label{XiDpropII}\\
  &\mathcal{H}_{\mathcal{D}}(\bm{\lambda})
  =\mathcal{H}_{\bar{\mathcal{D}}_{\II}}
  \bigl(\bm{\lambda}-(N+1)\tilde{\bm{\delta}}^{\II}\bigr),
  \label{HD=HDbII}\\
  &P_{\mathcal{D},n}(\eta;\bm{\lambda})\propto
  P_{\bar{\mathcal{D}}_{\II},n}\bigl(\eta;\bm{\lambda}
  -(N+1)\tilde{\bm{\delta}}^{\II}\bigr)\quad(n=0,1,\ldots).
  \label{PDnpropII}
\end{align}
We note that $\ell_{\mathcal{D}}=\ell_{\bar{\mathcal{D}}_{\I}}
=\ell_{\bar{\mathcal{D}}_{\II}}$ and
the energy eigenvalues are invariant under twisting shifts,
$\mathcal{E}_n(\bm{\lambda})
=\mathcal{E}_n(\bm{\lambda}+\tilde{\bm{\delta}}^{\I})
=\mathcal{E}_n(\bm{\lambda}+\tilde{\bm{\delta}}^{\II})$ and
\begin{equation}
  \mathcal{E}_n(\bm{\lambda})
  =\mathcal{E}_n\bigl(\bm{\lambda}-(N+1)\tilde{\bm{\delta}}^{\I}\bigr)
  =\mathcal{E}_n\bigl(\bm{\lambda}-(N+1)\tilde{\bm{\delta}}^{\II}\bigr).
  \label{En=EnI=EnII}
\end{equation}

\subsection{Jacobi}
\label{sec:equiv_J}

Next we consider Jacobi case. The argument presented in the previous
subsection applies also to this case.
{}From \eqref{Xiid3} we obtain the proportionality \eqref{XiDpropI}.
By using \eqref{la-(N+1)I}, \eqref{XiDpropI} and general formulas \eqref{UD}
and \eqref{phiDn}--\eqref{PD0=XiD}, we can show \eqref{psiDprop} and
\begin{align}
  \phi_{\mathcal{D}\,0}(x;\bm{\lambda})
  \propto\phi_{\bar{\mathcal{D}}_{\I}\,0}
  \bigl(x;\bm{\lambda}-(N+1)\tilde{\bm{\delta}}^{\I}\bigr),\quad
  U_{\mathcal{D}}(x;\bm{\lambda})=U_{\bar{\mathcal{D}}_{\I}}
  \bigl(x;\bm{\lambda}-(N+1)\tilde{\bm{\delta}}^{\I}\bigr).
  \label{phiD0,UD}
\end{align}
Therefore we obtain \eqref{HD=HDbI} and \eqref{PDnpropI}.
We have also \eqref{NII}--\eqref{En=EnI=EnII}.

\subsection{Laguerre}
\label{sec:equiv_L}

Lastly we consider Laguerre case.
The (pseudo)virtual states are (see Appendix)
\begin{equation*}
  \tilde{\phi}^{\I}_{\text{v}}(x;\bm{\lambda})
  =i^{-g}\phi_{\text{v}}(ix;\bm{\lambda}),\quad
  \tilde{\phi}^{\II}_{\text{v}}(x;\bm{\lambda})
  =\phi_{\text{v}}\bigl(x;\mathfrak{t}(\bm{\lambda})\bigl),\quad
  \tilde{\phi}^{\text{pvs}}_{\text{v}}(x;\bm{\lambda})
  =i^{g-1}\phi_{\text{v}}\bigl(ix;\mathfrak{t}(\bm{\lambda})\bigl),
\end{equation*}
and we have
\begin{equation*}
  \tilde{\phi}^{\text{pvs}}_{\text{v}}(ix;\bm{\lambda})
  \propto\tilde{\phi}^{\II}_{\text{v}}(x;\bm{\lambda}),\quad
  \tilde{\phi}^{\text{pvs}}_{\text{v}}
  \bigl(x;\mathfrak{t}(\bm{\lambda})\bigr)
  \propto\tilde{\phi}^{\I}_{\text{v}}(x;\bm{\lambda}).
\end{equation*}
The property of the Wronskian ($f_j=f_j(x)$)
\begin{equation*}
  \text{W}[f_1,f_2,\ldots,f_n]\bigl(g(x)\bigr)
  =\bigl(\tfrac{dg(x)}{dx}\bigr)^{-\frac12n(n-1)}
  \text{W}[F_1,F_2,\ldots,F_n](x),\ \ F_j(x)\eqdef f_j\bigl(g(x)\bigr),
\end{equation*}
gives $\text{W}[f_1,\ldots,f_n](ix)\propto
\text{W}[f_1(ix),\ldots,f_n(ix)](x)$.
By using these, \eqref{Xiid3} with the replacement $x\to ix$ leads to
\eqref{XiDpropI}, and \eqref{Xiid3} with the replacement
$\bm{\lambda}\to\mathfrak{t}(\bm{\lambda})$ leads to \eqref{XiDpropII}.

Once \eqref{XiDpropI} and \eqref{XiDpropII} are obtained, the remaining
task is the same as Jacobi case.
By using \eqref{la-(N+1)I}, \eqref{XiDpropI} and general formulas \eqref{UD}
and \eqref{phiDn}--\eqref{PD0=XiD}, we can show \eqref{psiDprop} and
\eqref{phiD0,UD}.
Therefore we obtain \eqref{HD=HDbI} and \eqref{PDnpropI}.
We have also \eqref{NII}--\eqref{En=EnI=EnII}.

\subsection{Zero level deletion}
\label{sec:equiv_zero}

If the degree of virtual state wavefunction for deletion is zero
($d_j=0$), an $M$-indexed orthogonal polynomial reduces to an $(M-1)$-indexed
orthogonal polynomial \cite{os25,os27}:
\begin{align}
  P_{\mathcal{D},n}(\eta;\bm{\lambda})\Bigm|_{d^{\I}_{M_{\I}}=0}
  &\propto P_{\mathcal{D}',n}(x;\bm{\lambda}+\tilde{\bm{\delta}}^{\I}),\n
  &\quad\mathcal{D}'=\{d^{\I}_1-1,\ldots,d^{\I}_{M_{\I}-1}-1,
  d^{\II}_1+1,\ldots,d^{\II}_{M_{\II}}+1\},
  \label{dM1=0}\\
  P_{\mathcal{D},n}(x;\bm{\lambda})\Bigm|_{d^{\II}_{M_{\II}}=0}
  &\propto P_{\mathcal{D}',n}(x;\bm{\lambda}+\tilde{\bm{\delta}}^{\II}),\n
  &\quad\mathcal{D}'=\{d^{\I}_1+1,\ldots,d^{\I}_{M_{\I}}+1,
  d^{\II}_1-1,\ldots,d^{\II}_{M_{\II}-1}-1\},
  \label{d'M2=0}
\end{align}
which are the consequences of shape-invariance.
They can be rederived by \eqref{PDnpropI} and \eqref{PDnpropII}.
By using \eqref{PDnpropI}, the following index sets give equivalent
multi-indexed orthogonal polynomials:
\begin{align*}
  &\ \{d^{\I}_1,\ldots,d^{\I}_{M_{\I}-1},0,d^{\II}_1,\ldots,
  d^{\II}_{d_{M_{\II}}}\}\text{ with }\bm{\lambda}\qquad
  \bigl(N\geq\max\{d^{\II}_1,\ldots,d^{\II}_{M_{\II}}\}\bigr)\\
  \leftrightarrow&
  \ \Bigl(\{0,\ldots,N\}\backslash\{N-d^{\II}_1,\ldots,
  N-d^{\II}_{M_{\II}}\}\Bigr)
  \cup\{N+1+d^{\I}_1,\ldots,N+1+d^{\I}_{M_{\I}-1},N+1\}\\
  &\quad\text{(all type $\I$)}\quad
  \text{ with }\bm{\lambda}-(N+1)\tilde{\bm{\delta}}^{\I}\\
  =\,&\ \Bigl(\{0,\ldots,N+1\}\backslash\bigl\{(N+1)-(d^{\II}_1+1),\ldots,
  (N+1)-(d^{\II}_{M_{\II}}+1)\bigr\}\Bigr)\\
  &\quad\cup
  \bigl\{(N+1)+1+(d^{\I}_1-1),\ldots,(N+1)+1+(d^{\I}_{M_{\I}-1}-1)\bigr\}
  \quad\text{(all type $\I$)}\\
  &\quad\text{ with }\bm{\lambda}+\tilde{\bm{\delta}}^{\I}
  -\bigl((N+1)+1\bigr)\tilde{\bm{\delta}}^{\I}\qquad
  \bigl(N+1\geq\max\{d^{\II}_1+1,\ldots,d^{\II}_{M_{\II}}+1\}\bigr)\\
  \leftrightarrow&
  \ \{d^{\I}_1-1,\ldots,d^{\I}_{M_{\I}-1}-1,d^{\II}_1+1,\ldots,
  d^{\II}_{d_{M_{\II}}}+1\}\text{ with }\bm{\lambda}+\tilde{\bm{\delta}}^{\I},
\end{align*}
which means \eqref{dM1=0}.
Similarly \eqref{d'M2=0} is derived by \eqref{PDnpropII}.

\section{Summary and Comments}
\label{summary}

Multi-indexed orthogonal polynomials of L, J, W and AW types are labeled
by an index set $\mathcal{D}$ but different index sets may give the same
multi-indexed orthogonal polynomials, $P_{\mathcal{D},n}(\eta;\bm{\lambda})
\propto P_{\mathcal{D}',n}(\eta;\bm{\lambda}')$.
Based on extensions of Wronskian/Casoratian identities obtained in
\cite{os29,os30}, we have shown equivalences of two deformed systems,
\eqref{HD=HDbI} and \eqref{HD=HDbII}.
One system has both type $\I$ and type $\II$ indices and the other has
type $\I$ indices only (or type $\II$ indices only).
These equivalences of two systems imply the proportionalities of
multi-indexed orthogonal polynomials, \eqref{PDnpropI} and
\eqref{PDnpropII}.
Redundant $N$-dependence in \eqref{PDnpropI} and \eqref{PDnpropII}
comes from shape-invariance.
Two deformed systems with index sets of the form \eqref{D} (with $d_j>0$)
are equivalent, when the corresponding two $\bar{\mathcal{D}}_{\I}$ \eqref{DbI}
(or $\bar{\mathcal{D}}_{\II}$ \eqref{DbII}) with the minimum $N$ are equal
as a set (and parameters $\bm{\lambda}$ are appropriately shifted).

For the index sets with type $\I$ indices only (or type $\II$ indices only),
are there equivalences among them? We conjecture that there are no
equivalences among them (for generic parameters).
There are accidental coincidences of the denominator polynomials for low
degree cases, for example,
\begin{align*}
  &\mathcal{D}=\{2\},\ \ \mathcal{D}'=\{1,2\},
  \ \ \Xi_{\mathcal{D}}(\eta;\bm{\lambda})
  \propto\Xi_{\mathcal{D}'}(\eta;\bm{\lambda}'),\\
  &\ \text{type $\I$ only}:\ \ \bm{\lambda}'=\left\{
  \begin{array}{ll}
  (-2-g,4-h)&:\text{J}\\
  (\frac52-\lambda_1,\frac52-\lambda_2,-\frac12-\lambda_3,-\frac12-\lambda_4)
  &:\text{W,\,AW}
  \end{array}\right.,\\
  &\text{type $\II$ only}:\ \ \bm{\lambda}'=\left\{
  \begin{array}{ll}
  (4-g,-2-h)&:\text{J}\\
  (-\frac12-\lambda_1,-\frac12-\lambda_2,\frac52-\lambda_3,\frac52-\lambda_4)
  &:\text{W,\,AW}
  \end{array}\right..
\end{align*}
However these proportionalities of the denominator polynomials do not mean
those of the multi-indexed orthogonal polynomials.
In fact, we have $P_{\mathcal{D},n}(\eta;\bm{\lambda})\propto
P_{\mathcal{D}',n}(\eta;\bm{\lambda}'-2\bm{\delta})$ for $n=0,1$ but
$P_{\mathcal{D},n}(\eta;\bm{\lambda})\not{\!\propto}
P_{\mathcal{D}',n}(\eta;\bm{\lambda}'-2\bm{\delta})$ for $n\geq 2$.

Corresponding to the three term recurrence relations for ordinary orthogonal
polynomials, $M$-indexed orthogonal polynomials satisfy $3+2M$ term
recurrence relations \cite{rrmiop}.
When we calculate multi-indexed orthogonal polynomials explicitly,
this $3+2M$ term recurrence relation provides us more effective calculation
method compared to their original definitions which are expressed in terms
of determinant.
If the $M$-indexed orthogonal polynomials
$P_{\mathcal{D},n}(\eta;\bm{\lambda})$ are equivalent to $M'$-indexed one
$P_{\mathcal{D}',n}(\eta;\bm{\lambda}')$ ($M'<M$), these
$P_{\mathcal{D},n}(\eta;\bm{\lambda})$ also satisfy $3+2M'$ term
recurrence relations.
For an index set $\mathcal{D}$ \eqref{D} with parameters $\bm{\lambda}$,
this happens in the following cases (0)--(\romannumeral4):
\begin{equation*}
\text{
(0) $d_j=0$ : $\mathcal{D}'$ is given by 
\eqref{dM1=0} or \eqref{d'M2=0} ($M'=M-1$).
 }
\end{equation*}
In the following we assume $d_j>0$ and set
$I_1\eqdef 2M_{\I}-d^{\text{max}}_{\I}-1$ and
$I_2\eqdef 2M_{\II}-d^{\text{max}}_{\II}-1$.
We present $\mathcal{D}'$ with the minimum $M'$.
For $M_{\I},M_{\II}>0$ cases, we have
\begin{align*}
  \text{(\romannumeral1)}\ I_1\leq 0<I_2\text{ or }0<I_1\leq I_2
  &:\ \mathcal{D}'=\bar{\mathcal{D}}_{\I}\bigl|_{N=d^{\text{max}}_{\II}}
  \text{ with }\bm{\lambda}-(d^{\text{max}}_{\II}+1)\tilde{\bm{\delta}}^{\I}
  \ \ (M'=M-I_2),\\
  \text{(\romannumeral2)}\ I_2\leq 0<I_1\text{ or }0<I_2\leq I_1
  &:\ \mathcal{D}'=\bar{\mathcal{D}}_{\II}\bigl|_{N=d^{\text{max}}_{\I}}
  \text{ with }\bm{\lambda}-(d^{\text{max}}_{\I}+1)\tilde{\bm{\delta}}^{\II}
  \ \ (M'=M-I_1).
\end{align*}
For $M_{\I}=0$ or $M_{\II}=0$ cases, let us assume $d_1<\cdots<d_M$ and
use the formulas in the opposite direction.
Then we have for $M_{\II}=0$,
\begin{align*}
  &\text{(\romannumeral3)}\ \exists\,j\text{ s.t. }2j-1-d_j>0 :\\
  &\qquad\mathcal{D}'=\{(d_{L+1}-N-1)^{\I},\ldots,(d_M-N-1)^{\I}\}\cup
  \{d^{\II}_1,\ldots,d^{\II}_{N+1-L}\},\ \ N=d_L,\\
  &\qquad\text{with }\bm{\lambda}+(d_L+1)\tilde{\bm{\delta}}^{\I}
  \ \ \bigl(M'=M-(2L-1-d_L)\bigr),
\end{align*}
where $L$ is $j$ giving the maximum value of $2j-1-d_j$ ($L$ may not be unique)
and $d^{\II}_i$'s are determined by
\begin{align*}
  \{0,1,\ldots,N\}\backslash\{N-d^{\II}_1,\ldots,N-d^{\II}_{N+1-L}\}
  =\{d_1,\ldots,d_L\}.
\end{align*}
The case $M_{\I}=0$ is similar,
\begin{align*}
  &\text{(\romannumeral4)}\ \exists\,j\text{ s.t. }2j-1-d_j>0 :\\
  &\qquad\mathcal{D}'=\{(d_{L+1}-N-1)^{\II},\ldots,(d_M-N-1)^{\II}\}\cup
  \{d^{\I}_1,\ldots,d^{\I}_{N+1-L}\},\ \ N=d_L,\\
  &\qquad\text{with }\bm{\lambda}+(d_L+1)\tilde{\bm{\delta}}^{\II}
  \ \ \bigl(M'=M-(2L-1-d_L)\bigr),
\end{align*}
where $L$ is same as (\romannumeral3) and $d^{\I}_i$'s are determined by
$\{0,1,\ldots,N\}\backslash\{N-d^{\I}_1,\ldots,N-d^{\I}_{N+1-L}\}
=\{d_1,\ldots,d_L\}$.
For illustration we present examples:
\begin{align*}
  \text{(\romannumeral1)}:\ \,&
  \mathcal{D}=\{1^{\I},4^{\I},1^{\II},2^{\II}\}\text{ with }\bm{\lambda}
  \leftrightarrow\mathcal{D}'=\{2^{\I},4^{\I},7^{\I}\}\text{ with }
  \bm{\lambda}-3\tilde{\bm{\delta}}^{\I},\\
  \text{(\romannumeral2)}:\ \,&
  \mathcal{D}=\{1^{\I},2^{\I},4^{\I},2^{\II},3^{\II}\}\text{ with }\bm{\lambda}
  \leftrightarrow\mathcal{D}'=\{1^{\II},4^{\II},7^{\II},8^{\II}\}\text{ with }
  \bm{\lambda}-5\tilde{\bm{\delta}}^{\II},\\
  \text{(\romannumeral3)}:\ \,&
  \mathcal{D}=\{1^{\I},2^{\I},3^{\I},4^{\I},5^{\I}\}\text{ with }\bm{\lambda}
  \leftrightarrow\mathcal{D}'=\{5^{\II}\}\text{ with }
  \bm{\lambda}+6\tilde{\bm{\delta}}^{\I},\\
  &\mathcal{D}=\{2^{\I},3^{\I},4^{\I},7^{\I}\}\text{ with }\bm{\lambda}
  \leftrightarrow\mathcal{D}'=\{2^{\I},3^{\II},4^{\II}\}\text{ with }
  \bm{\lambda}+5\tilde{\bm{\delta}}^{\I},\\
  &\mathcal{D}=\{1^{\I},2^{\I},5^{\I},6^{\I}\}\text{ with }\bm{\lambda}
  \leftrightarrow\mathcal{D}'=\{2^{\II},3^{\II},6^{\II}\}\text{ with }
  \bm{\lambda}+7\tilde{\bm{\delta}}^{\I}\\
  &\phantom{\mathcal{D}=\{1^{\I},2^{\I},5^{\I},6^{\I}\}
  \text{ with }\bm{\lambda}}
  \leftrightarrow\mathcal{D}'=\{2^{\I},3^{\I},2^{\II}\}\text{ with }
  \bm{\lambda}+3\tilde{\bm{\delta}}^{\I},\\
  \text{(\romannumeral4)}:\ \,&
  \mathcal{D}=\{1^{\II},3^{\II},4^{\II},5^{\II},8^{\II}\}
  \text{ with }\bm{\lambda}
  \leftrightarrow\mathcal{D}'=\{3^{\I},5^{\I},2^{\II}\}\text{ with }
  \bm{\lambda}+6\tilde{\bm{\delta}}^{\II}.
\end{align*}

\medskip

\noindent
{\bf Note added:}
After submitting the manuscript (arXiv:1309.2346[math-ph]),
we have noticed that the proportionalities \eqref{PDnpropI} and
\eqref{PDnpropII} can be directly derived by repeated use of \eqref{dM1=0}
and \eqref{d'M2=0}, which is an idea of K.\,Takemura (private communication,
\cite{takemura}).

\section*{Acknowledgements}
I thank R.\,Sasaki for useful discussion and reading of the manuscript,
and K.\,Takemura for useful discussion.
I am supported in part by Grant-in-Aid for Scientific Research
from the Ministry of Education, Culture, Sports, Science and Technology
(MEXT), No.25400395.

\bigskip
\appendix
\section{Various Data}
\label{sec:app}

For reader's convenience, we present various data of quantum mechanical
systems described by Laguerre(L), Jacobi(J), Wilson(W) and Askey-Wilson(AW)
polynomials \cite{os25,os27,os29,os30}.

\subsection{Original systems}
\label{sec:A_org}

The Hamiltonian and eigenfunctions have the following form
\begin{align}
  &\mathcal{H}(\bm{\lambda})
  =\mathcal{A}(\bm{\lambda})^{\dagger}\mathcal{A}(\bm{\lambda}),\\
  &\phi_n(x;\bm{\lambda})=\phi_0(x;\bm{\lambda})\check{P}_n(x;\bm{\lambda}),
  \quad\check{P}_n(x;\bm{\lambda})\eqdef P_n\bigl(\eta(x);\bm{\lambda}\bigr),
\end{align}
and the systems are shape-invariant
\begin{equation}
  \mathcal{A}(\bm{\lambda})\mathcal{A}(\bm{\lambda})^{\dagger}
  =\kappa\mathcal{A}(\bm{\lambda}+\bm{\delta})^{\dagger}
  \mathcal{A}(\bm{\lambda}+\bm{\delta})
  +\mathcal{E}_1(\bm{\lambda}),
\end{equation}
which implies the forward and backward shift relations
\begin{align}
  &\mathcal{A}(\bm{\lambda})\phi_n(x;\bm{\lambda})
  =f_n(\bm{\lambda})\phi_{n-1}(x;\bm{\lambda}+\bm{\delta}),\quad
  \mathcal{A}(\bm{\lambda})^{\dagger}\phi_{n-1}(x;\bm{\lambda}+\bm{\delta})
  =b_{n-1}(\bm{\lambda})\phi_n(x;\bm{\lambda}),\\
  &\mathcal{F}(\bm{\lambda})\check{P}_n(x;\bm{\lambda})
  =f_n(\bm{\lambda})\check{P}_{n-1}(x;\bm{\lambda}+\bm{\delta}),\quad
  \mathcal{B}(\bm{\lambda})\check{P}_{n-1}(x;\bm{\lambda}+\bm{\delta})
  =b_{n-1}(\bm{\lambda})\check{P}_n(x;\bm{\lambda}),\\
  &\mathcal{F}(\bm{\lambda})\eqdef
  \phi_0(x;\bm{\lambda}+\bm{\delta})^{-1}\!\circ
  \mathcal{A}(\bm{\lambda})\circ\phi_0(x;\bm{\lambda}),
  \ \mathcal{B}(\bm{\lambda})\eqdef
  \phi_0(x;\bm{\lambda})^{-1}\!\circ
  \mathcal{A}(\bm{\lambda})^{\dagger}
  \circ\phi_0(x;\bm{\lambda}+\bm{\delta}),
\end{align}
with $\mathcal{E}_n(\bm{\lambda})=f_n(\bm{\lambda})b_{n-1}(\bm{\lambda})$.

\subsubsection{Laguerre and Jacobi}
\label{sec:A_org_LJ}

Operators $\mathcal{H}(\bm{\lambda})$, $\mathcal{A}(\bm{\lambda})$,
$\mathcal{A}(\bm{\lambda})^{\dagger}$, $\mathcal{F}(\bm{\lambda})$
and $\mathcal{B}(\bm{\lambda})$ are
\begin{align}
  &\mathcal{H}(\bm{\lambda})
  =-\frac{d^2}{dx^2}+U(x;\bm{\lambda}),\quad
  U(x;\bm{\lambda})=\frac{\partial_x^2\phi_0(x;\bm{\lambda})}
  {\phi_0(x;\bm{\lambda})},\\
  &\mathcal{A}(\bm{\lambda})\eqdef
  \frac{d}{dx}-\partial_x\log\bigl|\phi_0(x;\bm{\lambda})\bigr|,\quad
  \mathcal{A}(\bm{\lambda})^{\dagger}\eqdef
  -\frac{d}{dx}-\partial_x\log\bigl|\phi_0(x;\bm{\lambda})\bigr|,\\
  &\mathcal{F}(\bm{\lambda})=\left\{
  \begin{array}{ll}
  2\frac{d}{d\eta}&:\text{L}\\[3pt]
  -4\frac{d}{d\eta}&:\text{J}
  \end{array}\right.,\quad
  \mathcal{B}(\bm{\lambda})=\left\{
  \begin{array}{ll}
  -2\eta\frac{d}{d\eta}-2(g+\frac12-\eta)&:\text{L}\\[3pt]
  (1-\eta^2)\frac{d}{d\eta}+h-g+(g+h+1)\eta&:\text{J}
  \end{array}\right..
\end{align}
Various data of these systems are:
\begin{align}
  \text{L}:\quad
  &0<x<\infty,\quad
  \bm{\lambda}=g,\quad\bm{\delta}=1,\quad\kappa=1,\quad
  g>\tfrac12,\n
  &U(x;\bm{\lambda})=x^2+\frac{g(g-1)}{x^2}-(1+2g),\quad
  \eta(x)=x^2,\n
  &\mathcal{E}_n(\bm{\lambda})=4n,\quad
  f_n(\bm{\lambda})=-2,\quad b_{n-1}(\bm{\lambda})=-2n,\n
  &\phi_0(x;\bm{\lambda})=e^{-\frac12x^2}x^g,\quad
  P_n(\eta;\bm{\lambda})=L_n^{(g-\frac12)}(\eta),\n
  &h_n(\bm{\lambda})=\frac{1}{2\,n!}\,\Gamma(n+g+\tfrac12),\\
  \text{J}:\quad
  &0<x<\tfrac{\pi}{2},\quad
  \bm{\lambda}=(g,h),\quad\bm{\delta}=(1,1),\quad\kappa=1,\quad
  g,h>\tfrac12,\n
  &U(x;\bm{\lambda})=\frac{g(g-1)}{\sin^2x}
  +\frac{h(h-1)}{\cos^2 x}-(g+h)^2,\quad
  \eta(x)=\cos 2x,\n
  &\mathcal{E}_n(\bm{\lambda})=4n(n+g+h),\quad
  f_n(\bm{\lambda})=-2(n+g+h),\quad b_{n-1}(\bm{\lambda})=-2n,\n
  &\phi_0(x;\bm{\lambda})=(\sin x)^g(\cos x)^h,\quad
  P_n(\eta;\bm{\lambda})=P_n^{(g-\frac12,h-\frac12)}(\eta),\n
  &h_n(\bm{\lambda})=\frac{\Gamma(n+g+\frac12)\Gamma(n+h+\frac12)}
  {2\,n!\,(2n+g+h)\Gamma(n+g+h)},
\end{align}
where $L^{(\alpha)}_n(\eta)$ and $P^{(\alpha,\beta)}_n(\eta)$ are the
Laguerre and Jacobi polynomials \cite{koeswart} respectively.

\subsubsection{Wilson and Askey-Wilson}
\label{sec:A_org_WAW}

Operators $\mathcal{H}(\bm{\lambda})$, $\mathcal{A}(\bm{\lambda})$,
$\mathcal{A}(\bm{\lambda})^{\dagger}$, $\mathcal{F}(\bm{\lambda})$
and $\mathcal{B}(\bm{\lambda})$ are
\begin{align}
  &\mathcal{H}(\bm{\lambda})=
  \sqrt{V(x;\bm{\lambda})}\,e^{\gamma p}\sqrt{V^*(x;\bm{\lambda})}
  +\!\sqrt{V^*(x;\bm{\lambda})}\,e^{-\gamma p}\sqrt{V(x;\bm{\lambda})}
  -V(x;\bm{\lambda})-V^*(x;\bm{\lambda}),\\
  &\mathcal{A}(\bm{\lambda})\eqdef
  i\bigl(e^{\frac{\gamma}{2}p}\sqrt{V^*(x; \bm{\lambda})}
  -e^{-\frac{\gamma}{2}p}\sqrt{V(x;\bm{\lambda})}\,\bigr),\n
  &\mathcal{A}(\bm{\lambda})^{\dagger}\eqdef
  -i\bigl(\sqrt{V(x;\bm{\lambda})}\,e^{\frac{\gamma}{2}p}
  -\sqrt{V^*(x;\bm{\lambda})}\,e^{-\frac{\gamma}{2}p}\bigr),\\
  &\mathcal{F}(\bm{\lambda})
  =i\varphi(x)^{-1}(e^{\frac{\gamma}{2}p}-e^{-\frac{\gamma}{2}p}),\quad
  \mathcal{B}(\bm{\lambda})
  =-i\bigl(V(x;\bm{\lambda})e^{\frac{\gamma}{2}p}
  -V^*(x;\bm{\lambda})e^{-\frac{\gamma}{2}p}\bigr)\varphi(x).
\end{align}
 Various data of these systems are:
\begin{align}
  \text{W}:\quad
  &0<x<\infty,\ \ \gamma=1,
  \ \ \bm{\lambda}=(a_1,a_2,a_3,a_4),
  \ \ \bm{\delta}=(\tfrac12,\tfrac12,\tfrac12,\tfrac12),
  \ \ \kappa=1,\n
  &V(x;\bm{\lambda})=\frac{\prod_{j=1}^4(a_j+ix)}{2ix(2ix+1)},\quad
  \quad\eta(x)=x^2,\ \ \varphi(x)=2x,\n
  &\mathcal{E}_n(\bm{\lambda})=n(n+b_1-1),
  \ \ b_1\eqdef a_1+a_2+a_3+a_4,\n
  &f_n(\bm{\lambda})=-n(n+b_1-1),\quad
  b_{n-1}(\bm{\lambda})=-1,\n
  &\phi_0(x;\bm{\lambda})=
  \sqrt{\frac{\prod_{j=1}^4\Gamma(a_j+ix)\Gamma(a_j-ix)}
  {\Gamma(2ix)\Gamma(-2ix)}},\n
  &\check{P}_n(x;\bm{\lambda})=P_n\bigl(\eta(x);\bm{\lambda}\bigr)
  =W_n\bigl(\eta(x);a_1,a_2,a_3,a_4\bigr)\n
  &\qquad=(a_1+a_2,a_1+a_3,a_1+a_4)_n
  \ {}_4F_3\Bigl(
  \genfrac{}{}{0pt}{}{-n,\,n+b_1-1,\,a_1+ix,\,a_1-ix}
  {a_1+a_2,\,a_1+a_3,\,a_1+a_4}\Bigm|1\Bigr),\n
  &h_n(\bm{\lambda})=\frac{2\pi n!\,(n+b_1-1)_n
  \prod_{1\leq i<j\leq 4}\Gamma(n+a_i+a_j)}
  {\Gamma(2n+b_1)},\\
  \text{AW}:\quad
  &0<x<\pi,\ \ \gamma=\log q,
  \ \ q^{\bm{\lambda}}=(a_1,a_2,a_3,a_4),
  \ \ \bm{\delta}=(\tfrac12,\tfrac12,\tfrac12,\tfrac12),
  \ \ \kappa=q^{-1},\n
  &V(x;\bm{\lambda})=\frac{\prod_{j=1}^4(1-a_je^{ix})}
  {(1-e^{2ix})(1-qe^{2ix})},\quad
  \quad\eta(x)=\cos x,\ \ \varphi(x)=2\sin x,\n
  &\mathcal{E}_n(\bm{\lambda})=(q^{-n}-1)(1-b_4q^{n-1}),
  \ \ b_4\eqdef a_1a_2a_3a_4,\n
  &f_n(\bm{\lambda})=q^{\frac{n}{2}}(q^{-n}-1)(1-b_4q^{n-1}),\quad
  b_{n-1}(\bm{\lambda})=q^{-\frac{n}{2}},\n
  &\phi_0(x;\bm{\lambda})=
  \sqrt{\frac{(e^{2ix},e^{-2ix};q)_{\infty}}
  {\prod_{j=1}^4(a_je^{ix},a_je^{-ix};q)_{\infty}}},\n
  &\check{P}_n(x;\bm{\lambda})=P_n\bigl(\eta(x);\bm{\lambda}\bigr)
  =p_n\bigl(\eta(x);a_1,a_2,a_3,a_4|q\bigr)\n
  &\qquad=a_1^{-n}(a_1a_2,a_1a_3,a_1a_4;q)_n
  \ {}_4\phi_3\Bigl(\genfrac{}{}{0pt}{}{q^{-n},\,b_4q^{n-1},\,
  a_1e^{ix},\,a_1e^{-ix}}{a_1a_2,\,a_1a_3,\,a_1a_4}\!\!\Bigm|\!q\,;q\Bigr),\n
  &h_n(\bm{\lambda})=
  \frac{2\pi(b_4q^{n-1};q)_n(b_4q^{2n};q)_{\infty}}
  {(q^{n+1};q)_{\infty}\prod_{1\leq i<j\leq 4}(a_ia_jq^n;q)_{\infty}},
\end{align}
where $W_n(\eta;a_1,a_2,a_3,a_4)$ and $p_n(\eta;a_1,a_2,a_3,a_4|q)$ are
the Wilson and the Askey-Wilson polynomials \cite{koeswart} and
$q^{\bm{\lambda}}$ stands for
$q^{(\lambda_1,\lambda_2,\ldots)}=(q^{\lambda_1},q^{\lambda_2},\ldots)$
and $0<q<1$.
The parameters are restricted by
\begin{equation}
  \{a_1^*,a_2^*,a_3^*,a_4^*\}=\{a_1,a_2,a_3,a_4\}\ \ (\text{as a set});\quad
  \text{W}:\ \text{Re}\,a_j>0,\quad
  \text{AW}:\ |a_j|<1.
  \label{rangeorg}
\end{equation}

\subsection{Virtual state wavefunctions}
\label{sec:A_vs}

We have two types of virtual states,
type $\I$ $\tilde{\phi}^{\I}_{\text{v}}(x;\bm{\lambda})$ and
type $\II$ $\tilde{\phi}^{\II}_{\text{v}}(x;\bm{\lambda})$
($\text{v}\in\mathbb{Z}_{\geq 0}$).

\subsubsection{Laguerre and Jacobi}
\label{sec:A_vs_LJ}

\begin{align}
  \text{L1}:&\quad
  \tilde{\phi}^{\I}_{\text{v}}(x;\bm{\lambda})
  \eqdef i^{-g}\phi_{\text{v}}\bigl(ix;\mathfrak{t}^{\I}(\bm{\lambda})\bigr),
  \ \ \xi^{\I}_{\text{v}}(\eta;\bm{\lambda})\eqdef
  P_{\text{v}}\bigl(-\eta;\mathfrak{t}^{\I}(\bm{\lambda})\bigr),\n
  &\quad\mathfrak{t}^{\I}\eqdef\text{id},
  \ \ \tilde{\bm{\delta}}^{\I}\eqdef 1,\quad
  \tilde{\mathcal{E}}^{\I}_{\text{v}}(\bm{\lambda})=-4(g+\text{v}+\tfrac12),\\
  \text{L2}:&\quad
  \tilde{\phi}^{\II}_{\text{v}}(x;\bm{\lambda})
  \eqdef\phi_{\text{v}}\bigl(x;\mathfrak{t}^{\II}(\bm{\lambda})\bigr),
  \ \ \xi^{\II}_{\text{v}}(\eta;\bm{\lambda})\eqdef
  P_{\text{v}}\bigl(\eta;\mathfrak{t}^{\II}(\bm{\lambda})\bigr),\n
  &\quad\mathfrak{t}^{\II}(\bm{\lambda})\eqdef 1-g,
  \ \ \tilde{\bm{\delta}}^{\II}\eqdef-1,\quad
  \tilde{\mathcal{E}}^{\II}_{\text{v}}(\bm{\lambda})=-4(g-\text{v}-\tfrac12),\\
  \text{J1}:&\quad
  \tilde{\phi}^{\I}_{\text{v}}(x;\bm{\lambda})
  \eqdef\phi_{\text{v}}\bigl(x;\mathfrak{t}^{\I}(\bm{\lambda})\bigr),
  \ \ \xi^{\I}_{\text{v}}(\eta;\bm{\lambda})\eqdef
  P_{\text{v}}\bigl(\eta;\mathfrak{t}^{\I}(\bm{\lambda})\bigr),\n
  &\quad\mathfrak{t}^{\I}(\bm{\lambda})\eqdef(g,1-h),
  \ \ \tilde{\bm{\delta}}^{\I}\eqdef(1,-1),\quad
  \tilde{\mathcal{E}}^{\I}_{\text{v}}(\bm{\lambda})
  =-4(g+\text{v}+\tfrac12)(h-\text{v}-\tfrac12),\\
  \text{J2}:&\quad
  \tilde{\phi}^{\II}_{\text{v}}(x;\bm{\lambda})
  \eqdef\phi_{\text{v}}\bigl(x;\mathfrak{t}^{\II}(\bm{\lambda})\bigr),
  \ \ \xi^{\II}_{\text{v}}(\eta;\bm{\lambda})\eqdef
  P_{\text{v}}\bigl(\eta;\mathfrak{t}^{\II}(\bm{\lambda})\bigr),\n
  &\quad\mathfrak{t}^{\II}(\bm{\lambda})\eqdef(1-g,h),
  \ \ \tilde{\bm{\delta}}^{\II}\eqdef(-1,1),\quad
  \tilde{\mathcal{E}}^{\II}_{\text{v}}(\bm{\lambda})
  =-4(g-\text{v}-\tfrac12)(h+\text{v}+\tfrac12).
\end{align}
We have changed the signs of $\tilde{\bm{\delta}}^{\I}$ and
$\tilde{\bm{\delta}}^{\II}$ from those in \cite{os25}.

\subsubsection{Wilson and Askey-Wilson}
\label{sec:A_vs_WAW}

\begin{align}
  \text{type $\I$}:&\quad
  \tilde{\phi}^{\I}_{\text{v}}(x;\bm{\lambda})
  \eqdef\phi_{\text{v}}\bigl(x;\mathfrak{t}^{\I}(\bm{\lambda})\bigr),
  \ \ \xi^{\I}_{\text{v}}(\eta;\bm{\lambda})\eqdef
  P_{\text{v}}\bigl(\eta;\mathfrak{t}^{\I}(\bm{\lambda})\bigr),
  \ \ \check{\xi}^{\I}_{\text{v}}(x;\bm{\lambda})\eqdef
  \xi^{\I}_{\text{v}}\bigl(\eta(x);\bm{\lambda}\bigr),\n
  &\quad
  \mathfrak{t}^{\I}(\bm{\lambda})
  \eqdef(1-\lambda_1,1-\lambda_2,\lambda_3,\lambda_4),
  \ \ \tilde{\bm{\delta}}^{\I}\eqdef(-\tfrac12,-\tfrac12,\tfrac12,\tfrac12),\\
  &\quad
  \tilde{\mathcal{E}}^{\I}_{\text{v}}(\bm{\lambda})=\left\{
  \begin{array}{ll}
  -(a_1+a_2-\text{v}-1)(a_3+a_4+\text{v})&\!\!:\text{W}\\
  -(1-a_1a_2q^{-\text{v}-1})(1-a_3a_4q^{\text{v}})&\!\!:\text{AW}
  \end{array}\right.\!\!,
  \ \ \alpha^{\I}(\bm{\lambda})=\left\{
  \begin{array}{ll}
  1&\!\!:\text{W}\\
  a_1a_2q^{-1}&\!\!:\text{AW}
  \end{array}\right.\!\!,\n
  \text{type $\II$}:&\quad
  \tilde{\phi}^{\II}_{\text{v}}(x;\bm{\lambda})
  \eqdef\phi_{\text{v}}\bigl(x;\mathfrak{t}^{\II}(\bm{\lambda})\bigr),
  \ \ \xi^{\II}_{\text{v}}(\eta;\bm{\lambda})\eqdef
  P_{\text{v}}\bigl(\eta;\mathfrak{t}^{\II}(\bm{\lambda})\bigr),
  \ \ \check{\xi}^{\II}_{\text{v}}(x;\bm{\lambda})\eqdef
  \xi^{\II}_{\text{v}}\bigl(\eta(x);\bm{\lambda}\bigr),\n
  &\quad
  \mathfrak{t}^{\II}(\bm{\lambda})
  \eqdef(\lambda_1,\lambda_2,1-\lambda_3,1-\lambda_4),
  \ \ \tilde{\bm{\delta}}^{\II}\eqdef(\tfrac12,\tfrac12,-\tfrac12,-\tfrac12),\\
  &\quad
  \tilde{\mathcal{E}}^{\II}_{\text{v}}(\bm{\lambda})=\left\{
  \begin{array}{ll}
  -(a_3+a_4-\text{v}-1)(a_1+a_2+\text{v})&\!\!:\text{W}\\
  -(1-a_3a_4q^{-\text{v}-1})(1-a_1a_2q^{\text{v}})&\!\!:\text{AW}
  \end{array}\right.\!\!,
  \ \ \alpha^{\II}(\bm{\lambda})=\left\{
  \begin{array}{ll}
  1&\!\!:\text{W}\\
  a_3a_4q^{-1}&\!\!:\text{AW}
  \end{array}\right.\!\!.\nonumber
\end{align}

\subsection{Pseudo virtual state wavefunctions}
\label{sec:A_pvs}

The twist operation $\mathfrak{t}$ is defined by
\begin{equation}
  \mathfrak{t}\eqdef\mathfrak{t}^{\II}\circ\mathfrak{t}^{\I}
  \ \ \bigl(\Rightarrow\ \mathfrak{t}=\mathfrak{t}^{\I}
  \circ\mathfrak{t}^{\II}\bigr),
  \label{t}
\end{equation}
and the energy of pseudo virtual state
$\tilde{\phi}^{\text{pvs}}_{\text{v}}(x;\bm{\lambda})$
($\text{v}\in\mathbb{Z}_{\geq 0}$) is
\begin{equation}
  \tilde{\mathcal{E}}^{\text{pvs}}_{\text{v}}(\bm{\lambda})
  =\mathcal{E}_{-\text{v}-1}(\bm{\lambda}).
\end{equation}

\subsubsection{Laguerre and Jacobi}
\label{sec:A_pvs_LJ}

\begin{align}
  \text{L}:&\quad
  \tilde{\phi}^{\text{pvs}}_{\text{v}}(x;\bm{\lambda})
  \eqdef i^{g-1}\phi_{\text{v}}\bigl(ix;\mathfrak{t}(\bm{\lambda})\bigr),
  \ \ \xi^{\text{pvs}}_{\text{v}}(\eta;\bm{\lambda})\eqdef
  P_{\text{v}}\bigl(-\eta;\mathfrak{t}(\bm{\lambda})\bigr),
  \ \ \mathfrak{t}(\bm{\lambda})=1-g,\\
  \text{J}:&\quad
  \tilde{\phi}^{\text{pvs}}_{\text{v}}(x;\bm{\lambda})
  \eqdef\phi_{\text{v}}\bigl(x;\mathfrak{t}(\bm{\lambda})\bigr),
  \ \ \xi^{\text{pvs}}_{\text{v}}(\eta;\bm{\lambda})\eqdef
  P_{\text{v}}\bigl(\eta;\mathfrak{t}(\bm{\lambda})\bigr),
  \ \ \mathfrak{t}(\bm{\lambda})=(1-g,1-h).
\end{align}

\subsubsection{Wilson and Askey-Wilson}
\label{sec:A_pvs_WAW}

\begin{align}
  &\tilde{\phi}^{\text{pvs}}_{\text{v}}(x;\bm{\lambda})
  \eqdef\phi_{\text{v}}\bigl(x;\mathfrak{t}(\bm{\lambda})\bigr),
  \ \ \xi^{\text{pvs}}_{\text{v}}(\eta;\bm{\lambda})\eqdef
  P_{\text{v}}\bigl(\eta;\mathfrak{t}(\bm{\lambda})\bigr),\n
  &\mathfrak{t}(\bm{\lambda})
  =(1-\lambda_1,1-\lambda_2,1-\lambda_3,1-\lambda_4).
\end{align}

\subsection{Multi-indexed orthogonal polynomials}
\label{sec:A_miop}

The set $\mathcal{D}$ is given in \eqref{D}.
The eigenfunctions $\phi_{\mathcal{D}\,n}(x;\bm{\lambda})$ of the
deformed system $\mathcal{H}_{\mathcal{D}}(\bm{\lambda})$ have the form
\eqref{phiDn}--\eqref{laI_II}.
The denominator polynomial $\Xi_{\mathcal{D}}(\eta;\bm{\lambda})$ and
the multi-indexed orthogonal polynomial $P_{\mathcal{D},n}(\eta;\bm{\lambda})$
are polynomials in $\eta$ and their degree are generically
$\ell_{\mathcal{D}}$ and $\ell_{\mathcal{D}}+n$, respectively
($\ell_{\mathcal{D}}$ is given by \eqref{ellD}).

\subsubsection{Laguerre and Jacobi}
\label{sec:A_miop_LJ}

\begin{align}
  &\Xi_{\mathcal{D}}(\eta;\bm{\lambda})\eqdef
  \text{W}[\mu_1,\ldots,\mu_{M_{\I}},\nu_1,\ldots,\nu_{M_{\II}}](\eta)\n
  &\phantom{\Xi_{\mathcal{D}}(\eta;\bm{\lambda})\eqdef}
  \quad\times\left\{
  \begin{array}{ll}
  e^{-M_{\I}\eta}\,\eta^{(M_{\I}+g-\frac12)M_{\II}}&:\text{L}\\[2pt]
  \bigl(\frac{1-\eta}{2}\bigr)^{(M_{\I}+g-\frac12)M_{\II}}
  \bigl(\frac{1+\eta}{2}\bigr)^{(M_{\II}+h-\frac12)M_{\I}}&:\text{J}
  \end{array}\right.,\\
  &P_{\mathcal{D},n}(\eta;\bm{\lambda})\eqdef
  \text{W}[\mu_1,\ldots,\mu_{M_{\I}},\nu_1,\ldots,\nu_{M_{\II}},P_n](\eta)\n
  &\phantom{P_{\mathcal{D},n}(\eta;\bm{\lambda})\eqdef}
  \quad\times\left\{
  \begin{array}{ll}
  e^{-M_{\I}\eta}\,\eta^{(M_{\I}+g+\frac12)M_{\II}}&:\text{L}\\[2pt]
  \bigl(\frac{1-\eta}{2}\bigr)^{(M_{\I}+g+\frac12)M_{\II}}
  \bigl(\frac{1+\eta}{2}\bigr)^{(M_{\II}+h+\frac12)M_{\I}}&:\text{J}
  \end{array}\right.,
  \label{miopLJ}\\
  &\mu_j=\xi_{d_j^{\I}}^{\I}(\eta;\bm{\lambda})\times\left\{
  \begin{array}{ll}
  e^{\eta}&:\text{L}\\
  \bigl(\frac{1+\eta}{2}\bigr)^{\frac12-h}&:\text{J}
  \end{array}\right.,\quad
  \nu_j=\xi_{d_j^{\II}}^{\II}(\eta;\bm{\lambda})\times\left\{
  \begin{array}{ll}
  \eta^{\frac12-g}&:\text{L}\\
  \bigl(\frac{1-\eta}{2}\bigr)^{\frac12-g}&:\text{J}
  \end{array}\right..
\end{align}

\subsubsection{Wilson and Askey-Wilson}
\label{sec:A_miop_WAW}

\begin{align}
  &\check{\Xi}_{\mathcal{D}}(x;\bm{\lambda})\eqdef
  \Xi_{\mathcal{D}}\bigl(\eta(x);\bm{\lambda}\bigr),\quad
  \check{P}_{\mathcal{D},n}(x;\bm{\lambda})\eqdef
  P_{\mathcal{D},n}\bigl(\eta(x);\bm{\lambda}\bigr),\\
  &\check{\Xi}_{\mathcal{D}}(x;\bm{\lambda})\eqdef
  A^{-1}\varphi_M(x)^{-1}\,i^{\frac12M(M-1)}\left|
  \begin{array}{llllll}
  \vec{X}^{(M)}_{d^{\I}_1}&\cdots&\vec{X}^{(M)}_{d^{\I}_{M_{\I}}}&
  \vec{Y}^{(M)}_{d^{\II}_1}&\cdots&\vec{Y}^{(M)}_{d^{\II}_{M_{\II}}}\\
  \end{array}\right|,\n
  &\qquad A=\left\{
  \begin{array}{ll}
  \prod_{k=3,4}\prod_{j=1}^{M_{\I}-1}
  (a_k-\frac{M-1}{2}+ix,a_k-\frac{M-1}{2}-ix)_j\\[4pt]
  \ \ \times\prod_{k=1,2}\prod_{j=1}^{M_{\II}-1}
  (a_k-\frac{M-1}{2}+ix,a_k-\frac{M-1}{2}-ix)_j
  &:\text{W}\\[4pt]
  \prod_{k=3,4}\prod_{j=1}^{M_{\I}-1}a_k^{-j}q^{\frac14j(j+1)}
  (a_kq^{-\frac{M-1}{2}}e^{ix},a_kq^{-\frac{M-1}{2}}e^{-ix};q)_j\\[4pt]
  \ \ \times\prod_{k=1,2}\prod_{j=1}^{M_{\II}-1}a_k^{-j}q^{\frac14j(j+1)}
  (a_kq^{-\frac{M-1}{2}}e^{ix},a_kq^{-\frac{M-1}{2}}e^{-ix};q)_j
  &:\text{AW}
  \end{array}\right.,
  \label{cXiDdef}\\[2pt]
  &\check{P}_{\mathcal{D},n}(x;\bm{\lambda})\eqdef
  B^{-1}\varphi_{M+1}(x)^{-1}\n
  &\phantom{\check{P}_{\mathcal{D},n}(x;\bm{\lambda})\eqdef}
  \times i^{\frac12M(M+1)}\left|
  \begin{array}{lllllll}
  \vec{X}^{(M+1)}_{d^{\I}_1}&\cdots&\vec{X}^{(M+1)}_{d^{\I}_{M_{\I}}}&
  \vec{Y}^{(M+1)}_{d^{\II}_1}&\cdots&\vec{Y}^{(M+1)}_{d^{\II}_{M_{\II}}}&
  \vec{Z}^{(M+1)}_n\\
  \end{array}\right|,\n
  &\qquad B=\left\{
  \begin{array}{ll}
  \prod_{k=3,4}\prod_{j=1}^{M_{\I}}
  (a_k-\frac{M}{2}+ix,a_k-\frac{M}{2}-ix)_j\\[4pt]
  \ \ \times\prod_{k=1,2}\prod_{j=1}^{M_{\II}}
  (a_k-\frac{M}{2}+ix,a_k-\frac{M}{2}-ix)_j
  &:\text{W}\\[4pt]
  \prod_{k=3,4}\prod_{j=1}^{M_{\I}}a_k^{-j}q^{\frac14j(j+1)}
  (a_kq^{-\frac{M}{2}}e^{ix},a_kq^{-\frac{M}{2}}e^{-ix};q)_j\\[4pt]
  \ \ \times\prod_{k=1,2}\prod_{j=1}^{M_{\II}}a_k^{-j}q^{\frac14j(j+1)}
  (a_kq^{-\frac{M}{2}}e^{ix},a_kq^{-\frac{M}{2}}e^{-ix};q)_j
  &:\text{AW}
  \end{array}\right.,
  \label{cPDndef}
\end{align}
where
\begin{align}
  &\bigl(\vec{X}^{(M)}_{\text{v}}\bigr)_j
  =r^{\II}_j(x^{(M)}_j;\bm{\lambda},M)
  \check{\xi}^{\I}_{\text{v}}(x^{(M)}_j;\bm{\lambda}),\qquad
  (1\leq j\leq M),\n
  &\bigl(\vec{Y}^{(M)}_{\text{v}}\bigr)_j
  =r^{\I}_j(x^{(M)}_j;\bm{\lambda},M)
  \check{\xi}^{\II}_{\text{v}}(x^{(M)}_j;\bm{\lambda}),\n
  &\bigl(\vec{Z}^{(M)}_n\bigr)_j
  =r^{\II}_j(x^{(M)}_j;\bm{\lambda},M)r^{\I}_j(x^{(M)}_j;\bm{\lambda},M)
  \check{P}_n(x^{(M)}_j;\bm{\lambda}),
\end{align}
and $x_j^{(n)}=x+i(\tfrac{n+1}{2}-j)\gamma$ and
\begin{align}
  r^{\I}_j(x^{(M)}_j;\bm{\lambda},M)
  &=\alpha^{\I}\bigl(\bm{\lambda}
  +(M-1)\tilde{\bm{\delta}}^{\I}\bigr)^{-\frac12(M-1)}
  \kappa^{\frac12(M-1)^2-(j-1)(M-j)}\\
  &\quad\times\left\{
  \begin{array}{ll}
  {\displaystyle
  \prod_{k=1,2}(a_k-\tfrac{M-1}{2}+ix)_{j-1}(a_k-\tfrac{M-1}{2}-ix)_{M-j}
  }&:\text{W}\\
  {\displaystyle
  e^{ix(M+1-2j)}\prod_{k=1,2}(a_kq^{-\frac{M-1}{2}}e^{ix};q)_{j-1}
  (a_kq^{-\frac{M-1}{2}}e^{-ix};q)_{M-j}
  }&:\text{AW}
  \end{array}\right.,\n
  r^{\II}_j(x^{(M)}_j;\bm{\lambda},M)
  &=\alpha^{\II}\bigl(\bm{\lambda}
  +(M-1)\tilde{\bm{\delta}}^{\II}\bigr)^{-\frac12(M-1)}
  \kappa^{\frac12(M-1)^2-(j-1)(M-j)}\\
  &\quad\times\left\{
  \begin{array}{ll}
  {\displaystyle
  \prod_{k=3,4}(a_k-\tfrac{M-1}{2}+ix)_{j-1}(a_k-\tfrac{M-1}{2}-ix)_{M-j}
  }&:\text{W}\\
  {\displaystyle
  e^{ix(M+1-2j)}\prod_{k=3,4}(a_kq^{-\frac{M-1}{2}}e^{ix};q)_{j-1}
  (a_kq^{-\frac{M-1}{2}}e^{-ix};q)_{M-j}
  }&:\text{AW}
  \end{array}\right..\nonumber
\end{align}
The auxiliary function $\varphi_M(x)$ is defined by
\begin{equation}
  \varphi_M(x)\eqdef
  \varphi(x)^{[\frac{M}{2}]}\prod_{k=1}^{M-2}
  \bigl(\varphi(x-i\tfrac{k}{2}\gamma)\varphi(x+i\tfrac{k}{2}\gamma)
  \bigr)^{[\frac{M-k}{2}]},
\end{equation}
and $\varphi_0(x)=\varphi_1(x)=1$, see \cite{gos}.
Here $[x]$ denotes the greatest integer not exceeding $x$.

\subsubsection{coefficients of the highest degree terms}
\label{sec:A_coeff}

The coefficients of the highest degree term of the polynomials
$\Xi_{\mathcal{D}}$ and $P_{\mathcal{D},n}$ are
\begin{align}
  &\Xi_{\mathcal{D}}(\eta;\bm{\lambda})
  =c_{\mathcal{D}}^{\Xi}(\bm{\lambda})\eta^{\ell_{\mathcal{D}}}
  +(\text{lower order terms}),\n
  &P_{\mathcal{D}}(\eta;\bm{\lambda})
  =c_{\mathcal{D},n}^{P}(\bm{\lambda})\eta^{\ell_{\mathcal{D}}+n}
  +(\text{lower order terms}),
\end{align}
\begin{align}
  c_{\mathcal{D}}^{\Xi}(\bm{\lambda})&=
  \prod_{j=1}^{M_{\I}}c^{\I}_{d^{\I}_j}(\bm{\lambda})\cdot
  \prod_{j=1}^{M_{\II}}c^{\II}_{d^{\II}_j}(\bm{\lambda})\n
  &\quad\times\left\{
  \begin{array}{ll}
  \prod\limits_{1\leq j<k\leq M_{\I}}(d^{\I}_k-d^{\I}_j)\cdot
  \prod\limits_{1\leq j<k\leq M_{\II}}(d^{\II}_k-d^{\II}_j)
  &:\text{L,\,J,\,W}\\[12pt]
  \prod\limits_{1\leq j<k\leq M_{\I}}
  \tfrac12q^{\frac12(d^{\I}_j-d^{\I}_k)}(1-q^{d^{\I}_k-d^{\I}_j})\cdot
  \prod\limits_{1\leq j<k\leq M_{\II}}
  \tfrac12q^{\frac12(d^{\II}_j-d^{\II}_k)}(1-q^{d^{\II}_k-d^{\II}_j})
  &:\text{AW}
  \end{array}\right.\n[4pt]
  &\quad\times\left\{
  \begin{array}{ll}
  (-1)^{M_{\I}M_{\II}}
  &:\text{L}\\[2pt]
  \prod_{j=1}^{M_{\I}}\prod_{k=1}^{M_{\II}}
  \frac14(g-h+d^{\I}_j-d^{\II}_k)
  &:\text{J}\\[4pt]
  \prod_{j=1}^{M_{\I}}\prod_{k=1}^{M_{\II}}
  (-a_3-a_4-d^{\I}_j+a_1+a_2+d^{\II}_k)
  &:\text{W}\\[4pt]
  \prod_{j=1}^{M_{\I}}\prod_{k=1}^{M_{\II}}\frac{2}{\sqrt{a_1a_2a_3a_4}}\,
  q^{j+k-2-\frac12(d^{\I}_j+d^{\II}_k)}
  (a_3a_4q^{d^{\I}_j}-a_1a_2q^{d^{\II}_k})
  &:\text{AW}
  \end{array}\right.,
  \label{cXiD}\\
  c_{\mathcal{D},n}^{P}(\bm{\lambda})&=
  c_{\mathcal{D}}^{\Xi}(\bm{\lambda})c_n(\bm{\lambda})\n
  &\quad\times\left\{
  \begin{array}{ll}
  (-1)^{M_{\I}}\prod_{j=1}^{M_{\II}}(g+n-d^{\II}_j-\frac12)
  &:\text{L}\\[4pt]
  \prod_{j=1}^{M_{\I}}\frac12(h+n-d^{\I}_j-\frac12)\cdot
  \prod_{j=1}^{M_{\II}}\frac{-1}{2}(g+n-d^{\II}_j-\frac12)
  &:\text{J}\\[6pt]
  \prod_{j=1}^{M_{\I}}(-a_1-a_2-n+d^{\I}_j+1)\cdot
  \prod_{j=1}^{M_{\II}}(-a_3-a_4-n+d^{\II}_j+1)
  &:\text{W}\\[6pt]
  q^{2M_{\I}M_{\II}}
  \prod_{j=1}^{M_{\I}}(a_1a_2)^{-\frac12}q^{\frac12(d^{\I}_j+1-n)}
  (1-a_1a_2q^{n-d^{\I}_j-1})\\[4pt]
  \quad\times
  \prod_{j=1}^{M_{\II}}(a_3a_4)^{-\frac12}q^{\frac12(d^{\II}_j+1-n)}
  (1-a_3a_4q^{n-d^{\II}_j-1})
  &:\text{AW}
  \end{array}\right.\!\!\!\!,\!\!
  \label{cPDn}
\end{align}
where $c_n$, $c^{\I}_n$ and $c^{\II}_n$ are
\begin{align}
  &P_n(\eta;\bm{\lambda})=c_n(\bm{\lambda})\eta^n
  +(\text{lower order terms}),\\
  &c_n(\bm{\lambda})=\left\{
  \begin{array}{ll}
  {\displaystyle\frac{(-1)^n}{n!}}&:\text{L}\\[5pt]
  {\displaystyle\frac{(n+g+h)_n}{2^n\,n!}}&:\text{J}
  \end{array}\right.,\quad
  c_n(\bm{\lambda})=\left\{
  \begin{array}{ll}
  (-1)^n(n+b_1-1)_n&:\text{W}\\[2pt]
  2^n(b_4q^{n-1};q)_n&:\text{AW}
  \end{array}\right.,\\
  &c^{\I}_{\text{v}}(\bm{\lambda})\eqdef\left\{
  \begin{array}{ll}
  (-1)^{\text{v}}c_{\text{v}}(\bm{\lambda})&:\text{L}\\[1pt]
  c_{\text{v}}\bigl(\mathfrak{t}^{\I}(\bm{\lambda})\bigr)&:\text{J,\,W,\,AW}
  \end{array}\right.,\quad
  c^{\II}_{\text{v}}(\bm{\lambda})\eqdef
  c_{\text{v}}\bigl(\mathfrak{t}^{\II}(\bm{\lambda})\bigr)
  \ :\text{L,\,J,\,W,\,AW}.
\end{align}


\end{document}